\documentclass{aa}  
\usepackage{psfig}
\def\bj{\hbox{$b_j$}} 
\def\Bj{\hbox{$b_j$}}
\def\fss{{\em Fornax Spectroscopic Survey}}
\def\FSS{{\em FSS}}
\def\ah{\hbox{$^{\rm h}$}}
\def\am{\hbox{$^{\rm m}$}}

\def\Mpc{{\rm\thinspace Mpc}}
\def\asec{{\rm\thinspace arcsec}}
\def\deg{{\rm\thinspace deg}}
\def\degsq{\hbox{$\deg^{2}\,$}}

\def\mag{{\rm\thinspace mag}}

\def\Bu{\hbox{$\thinspace\bj\mag\asec^{-2}\,$}}
\def\s{{\rm\thinspace s}}

\def\km{{\rm\thinspace km}}
\def\kms{\hbox{$\km\s^{-1}\,$}}

\def\mpc3{\hbox{$\Mpc^{-3}\,$}}

\def\one_wide{8.5cm}
\def\two_wide{14.0cm}
\def\caii{{\rm Ca\hbox{\small II}}}

\def\ha{{\rm H\hbox{$\alpha$}}}
\def\hb{{\rm H\hbox{$\beta$}}}
\def\hd{{\rm H\hbox{$\delta$}}}
\def\he{{\rm H\hbox{$\epsilon$}}}
\def\hii{H\,{\small II}}
\def\nii{{\rm N\hbox{\small II}}}

\def\oiii{{\rm O\hbox{\small III}}}
\begin{document}

   \thesaurus{04     
              (05.01.1;   
               11.01.2;   
               11.19.7;   
               08.19.1;   
               04.19.1;   
               03.20.8)}  

   \title{The Fornax Spectroscopic Survey}

   \subtitle{I. Survey Strategy and Preliminary Results on the Redshift 
    Distribution of a Complete Sample of Stars and Galaxies}

   \author{
       M.J. Drinkwater\inst{1}, 
       S. Phillipps\inst{2},
       J.B. Jones\inst{2},
       M.D. Gregg\inst{3},
       J.H. Deady\inst{2},
       J.I. Davies\inst{4},
       Q.A. Parker\inst{5},
       E.M. Sadler\inst{6}, 
       R.M. Smith\inst{4} }

   \offprints{M.J. Drinkwater}

   \institute{
       School of Physics, University of Melbourne, 
       Victoria 3010, Australia
     \and
       Department of Physics, University of Bristol, 
       Tyndall Avenue, Bristol, BS8 1TL, England, U.K. 
     \and
       University of California, Davis, 
       and Institute for Geophysics and Planetary Physics, 
       Lawrence Livermore National Laboratory,
       L-413, Livermore, CA 94550, USA
     \and
       Department of Physics and Astronomy, University of Wales Cardiff,
       P.O. Box 913, Cardiff, CF2 3YB, Wales, U.K.
     \and
       Royal Observatory Edinburgh, Blackford Hill, Edinburgh, 
       EH9 3HJ, Scotland, U.K.
     \and
       School of Physics, University of Sydney, NSW 2006, Australia
     }

   \date{Received ; accepted }

   \titlerunning{Fornax Spectroscopic Survey}

   \authorrunning{M.J. Drinkwater et al.}

   \maketitle

%
%

\begin{abstract}
The {\em Fornax Spectroscopic Survey} will use the Two degree Field
spectrograph (2dF) of the Anglo-Australian Telescope to obtain spectra
for a complete sample of all 14000 objects with $16.5\leq\Bj\leq19.7$
in a 12 square degree area centred on the Fornax Cluster. The aims of
this project include the study of dwarf galaxies in the cluster (both
known low surface brightness objects and putative normal surface
brightness dwarfs) and a comparison sample of background field
galaxies. We will also measure quasars and other active galaxies, any
previously unrecognised compact galaxies and a large sample of
Galactic stars. By selecting all objects---both stars and
galaxies---independent of morphology, we cover a much larger range of
surface brightness and scale size than previous surveys.

In this paper we first describe the design of the survey. Our targets 
are selected from UK Schmidt Telescope sky survey plates digitised by 
the Automated Plate Measuring (APM) facility. We then describe the 
photometric and astrometric calibration of these data and show that 
the APM astrometry is accurate enough for use with the 2dF. We also 
describe a general approach to object identification using 
cross-correlations which allows us to identify and classify both 
stellar and galaxy spectra. 

We present results from the first 2dF field. Redshift distributions
and velocity structures are shown for all observed objects in the
direction of Fornax, including Galactic stars, galaxies in and around
the Fornax Cluster, and for the background galaxy population.  The
velocity data for the stars show the contributions from the different
Galactic components, plus a small tail to high velocities.  We find no
galaxies in the foreground to the cluster in our 2dF field.  The
Fornax Cluster is clearly defined kinematically. The mean velocity
from the 26 cluster members having reliable redshifts is $1560 \pm 80
\:\mbox{km}\:\mbox{s}^{-1}$. They show a velocity dispersion of $380
\pm 50 \:\mbox{km}\:\mbox{s}^{-1}$.  Large-scale structure can be
traced behind the cluster to a redshift beyond $z=0.3$. Background compact
galaxies and low surface brightness galaxies are found to follow the
general galaxy distribution.
\end{abstract}


%
%

\section{Introduction}
\label{sec_intro}


The development of a new generation of multi-object spectrographs,
exemplified by the `Two degree Field', or 2dF, multi-fibre spectrograph
on the Anglo-Australian Telescope (AAT), has opened up whole new areas
of astronomical survey science. One particular area, which we discuss
in this paper, is the opportunity to make a truly complete
spectroscopic survey of a given area on the sky, down to well
determined, faint limits, irrespective of image morphology or any
other preselection of target type.

The {\em Fornax Spectroscopic Survey}, or \FSS, seeks to exploit
the huge multiplexing advantage of 2dF by surveying a region of 12
square degrees centred on the Fornax Cluster of galaxies. It will
encompass both cluster galaxies, of a wide range of types and
magnitudes, and background and foreground galaxies (over a similarly
wide range of morphologies), as well as Galactic stars, QSOs and 
any unusual or rare objects. 

Although many surveys of nearby clusters have been made over the past
20 years or more, these are all limited in several crucial
aspects. Spectroscopic surveys exist, but typically only of the few
dozen brightest cluster galaxies (and any background interlopers in
the top few magnitudes of the cluster luminosity
function). Photometric surveys, of course, go much deeper, but such
studies must be of a statistical nature (e.g.\ subtracting off the
expected background numbers; Smith et al.\ 1997), or rely on
subjective judgements of likely cluster membership based on
morphology, surface brightness or colour (e.g.\ Ferguson 1989).  Of
particular concern is the surface brightness; low surface brightness
galaxies (LSBGs) seen towards a cluster are conventionally assumed to
be members, while apparently faint, but high surface brightness
galaxies (HSBGs) are presumed to be luminous objects in the background
(e.g.\ Sandage, Binggeli, \& Tammann 1985).  The failure of either
assumption, i.e.\ the existence of large background LSBGs (such as the
serendipitously discovered Malin 1; Bothun et al.\ 1987) or of a
population of high surface brightness (compact) dwarfs in the cluster
(Drinkwater \& Gregg 1998), can have a dramatic effect on our
perception of the galaxy population as a whole. Furthermore, it is
possible that a population of extremely compact galaxies (either in
the cluster or beyond) could masquerade as stars and hence be missed
altogether from galaxy samples. Examples have previously been found
in, for example, QSO surveys, but again these are serendipitous
discoveries and hard to quantify (see Drinkwater et al.\ 1999a = Paper~II, 
and references therein).

Few previous attempts at all-object surveys have been made. The one
most similar to ours is probably that of Morton and Tritton in the
early 1980s. They obtained around 600 objective prism spectra and
100 slit spectra for objects in a 0.31
square degree region of background sky (i.e.\ no prominent cluster)
over the course of a 5 year period (Morton, Krug \& Tritton 
1985). More recently Colless et al.\ (1993) obtained spectra of 
about 100 objects in a small area of sky and small magnitude
range in order to investigate the completeness of faint galaxy
redshift surveys.

Our overall survey will therefore represent a huge increase in the
volume of data and in addition will give a uniquely complete picture
of a cluster of galaxies.  It is worth noting that the huge galaxy
surveys planned, with 2dF (Folkes et al. 1999; Colless 1999) 
or the Sloan Digital Survey
(Gunn 1995; Loveday \& Pier 1998) will not address such
problems, since their galaxy samples will be pre-selected from
photometric surveys and will only include objects classified as
galaxies and not of too low surface brightness, thus removing both
ends of any potentially wide range of galaxy parameters.

In the present paper we discuss the design and aims of our all-object
{\em Fornax Spectroscopic Survey} and present initial results on the 
velocity distributions. Section~\ref{sec_survey} gives a
technical definition of the survey, describing the relevant features
of the 2dF spectrograph, the selection of our target catalogue and the
calibration of this input catalogue. In Section~\ref{sec_science} we
discuss the scientific aims of the survey and summarise the types and
numbers of objects we expect to observe. In Section~\ref{sec_obs} we
discuss the spectroscopic observations and observational strategy. We
describe the technique we have developed to identify and classify
objects automatically from the 2dF spectra and give some examples from
our initial observations. Section~\ref{sec_initsci} gives the initial 
redshift results and Section~\ref{sec_summary} summarises the survey 
work to date. 


\section{The Survey Design}
\label{sec_survey}

In this Section we describe the basic parameters of the \fss. We start
with the relevant technical details of the 2dF spectrograph, and then
discuss our selection of targets from the digitised photographic sky
survey plates and the calibration of our input catalogues.

\subsection{The 2dF Spectrograph}


The 2dF facility (Taylor, Cannon \& Parker 1998) is probably the most
complex ground-based astronomical instrument built to date.  Via a
straightforward `top-end' change the capability of the 3.9-m\
Anglo-Australian Telescope is transformed into an unique wide-field
multi-fibre spectroscopic survey instrument. Up to 400 fibres are
available at any one time for rapid configuration over the full
two-degree diameter focal surface via a highly accurate robotic
positioner mounted in situ. Each 2\asec\ diameter fibre can be
placed to an accuracy of 0.3\asec\ in less than 10 seconds. The
input target positions must be accurate to 0.3\asec\ r.m.s.\ or
better over the whole two-degree field to avoid vignetting of the fibre
entrance apertures. This requirement is only for relative positions;
the absolute accuracy of a complete set of targets and guide stars
need only be 1--2\asec\ as the guide stars will then centre all the
targets accurately.

The wide field is provided by a highly sophisticated multi-component
corrector with in-built atmospheric dispersion compensator. In a novel
arrangement 2dF can simultaneously observe 400 target objects on the
sky at the same time that a further 400 fibres are being configured
using the robotic positioner on one of the two available `field
plates' (focal surfaces). Once observations and configurations have
been completed (usually over the same timescale) a tumbling mechanism
allows the newly configured field plate to point at the sky whilst the
previously observed field can be re-configured for the next target
field. In this way rapid field inter-change is provided for an
extremely efficient observing environment. Each set of 400 fibres
feeds two spectrographs which accept 200 fibres each. These are
mounted on the 2dF top end ring and can produce low to medium
resolution spectra on the dedicated $1024\times1024$ TEK CCDs.
 
The 2dF is now operating at close to the original specifications
anticipated for this most complex of instruments (Lewis, Glazebrook \&
Taylor, 1998). Field configuration times of about 1 hour for 400
fibres permit rapid cycling of target fields and have enabled
excellent progress to be made with our complete survey.

\subsection{The Fornax Cluster Field}


We chose the Fornax Cluster for this study because it is a
well-studied, compact, nearby southern galaxy cluster suited to this
type of survey.  We and several other groups have made photometric or
small-scale spectroscopic surveys of the region (e.g.\ Ferguson 1989;
Davies et al.\ 1988; Drinkwater \& Gregg 1998, Hilker et al.\ 1999).
The published spectroscopic samples have either been very small 
or have concentrated on the brighter cluster galaxies (Jones \& Jones 1980; 
Drinkwater et al.\ 2000a). A search of NED\footnote{The NASA/IPAC
Extragalactic Database (NED) is operated by the Jet Propulsion
Laboratory, California Institute of Technology, under contract with
the National Aeronautics and Space Administration.} in our central 2dF
field (number 1 in Table~\ref{tab_fields}) found only 42 objects
brighter than $B=20$ with measured redshifts: 30 cluster galaxies, 6
background galaxies and 6 QSOs.  With 2dF we can now measure the
redshifts of some 700--900 galaxies and quasars in this same field.

\begin{figure}
\psfig{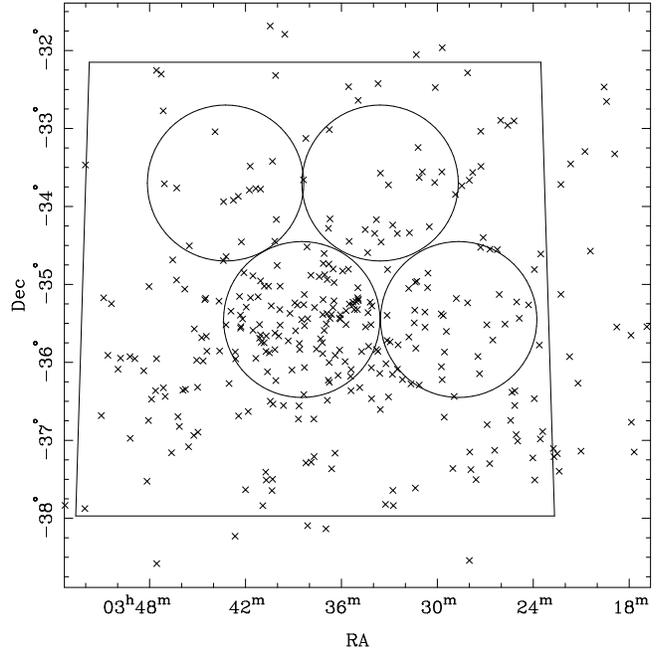}
\centering
\caption{
Distribution of the 2dF fields observed on the sky. The crosses indicate
the positions of the 340 members listed in the Fornax Cluster Catalog
(Ferguson 1989). The large trapezium shows the boundary of the region 
imaged on the Schmidt plate. 
}
\label{fig_sky}
\end{figure}

The Fornax Cluster is concentrated within one United Kingdom Schmidt
Telescope (UKST) Sky Survey plate and our survey will comprise four
separate 2dF fields which are listed in Table~\ref{tab_fields}. We
show the distribution of our fields on the sky in Fig.~\ref{fig_sky}
compared to the positions of galaxies classified as likely cluster
members by Ferguson (1989). Our first field is centred on the large
galaxy NGC~1399 at the centre of the cluster.  In order both to cover
a large number of targets and to go significantly deeper than previous
spectroscopic surveys we chose to limit our survey at a \Bj\ magnitude
of 19.7. This is then essentially the same depth as the large scale
2dF Galaxy Redshift Survey (GRS) of Ellis, Colless and collaborators
(e.g.\ Colless 1999; Folkes et al. 1999). This combination of survey
 area and magnitude limit will optimise our measurement of the cluster 
galaxies (see Section~\ref{sec_cluster}).

\begin{table}
\caption{Fornax Spectroscopic Survey Fields}
\label{tab_fields}
\begin{tabular}{lccl}
\hline 
N & RA (J2000) Dec & {\em l, b} & Comments\\
\hline 
  1  & 03 38 29.0 $-$35 27 01 & 236.7, $-$53.6 & NGC1399 \\
  2  & 03 28 40.0 $-$35 27 01 & 236.8, $-$55.6 \\
  3  & 03 33 38.0 $-$33 41 59 & 233.7, $-$54.6 \\
  4  & 03 43 15.0 $-$33 41 59 & 233.8, $-$52.6 \\
\hline 
\end{tabular}
\end{table}

\subsection{Target Selection}


In common with the large 2dF Galaxy Redshift Survey (Colless 1999), we
have chosen to select our targets from catalogues based on UKST Sky
Survey plates digitised by the Automatic Plate Measuring facility
(APM) at Cambridge. Unlike most galaxy surveys however, which only
select resolved images for spectroscopic measurement, we avoid any
morphological selection and include all objects, both resolved and
unresolved (i.e.\ `stars' and `galaxies').  This means that we can
include galaxies with the greatest possible range in surface
brightnesses (and with a range of scale sizes): 
our only selection criterion is a (blue) magnitude
limit. Including objects normally classified as stars greatly
increases the size of our sample but it is the only way to ensure
completeness.

Our input catalogue for the \FSS\ is a standard APM `catalogues' file
(see Irwin, Maddox \& McMahon 1994)
of field F358 from the UK Schmidt southern sky survey. The field is
centred at 03\ah37\am55\fs9, $-$34\degr50\arcmin14\arcsec\ (J2000) and
the region scanned for the catalogue file is $5.8\deg\times5.8\deg$.
This field is approximately centered on the Fornax Cluster and is
slightly smaller than the region surveyed by Ferguson (1989) shown in
Fig.~\ref{fig_sky}. The APM image catalogue lists image positions,
magnitudes and morphological classifications (as `star', `galaxy',
`noise', or `merged') measured from both the blue (\Bj) and red survey
plates.  The `merged' image classification indicates two overlapping
images: at the magnitudes of interest for this project the merged
objects nearly always consisted of a star overlapping a much fainter
galaxy. All the positions are measured from the more recent red survey
plate (epoch 1991 September 13 compared to 1976 November 18 for the
blue plate) to minimise problems with proper motions.  The APM
catalogue magnitudes are calibrated for unresolved (stellar) objects
only, so we supplemented these with total magnitudes for the galaxies
measured by direct analysis of the plate data (see
Section~\ref{sec_phot}).

Our target selection consisted simply of taking all objects from the
APM catalogue in each of our four 2dF fields with magnitudes in the
range $16.5\leq\Bj\leq19.7$. We did not apply any morphological selection,
although the APM image classifications from the blue survey plate were
used to determine which photometry to use (see Section~\ref{sec_phot}).
The limits were chosen to avoid very bright objects which could not be
observed efficiently with 2dF and for which the photographic
photometry would be unreliable, whilst, at the faint end, to allow us to
measure a significant area of the cluster (12\degsq) in a reasonable
amount of time.  With these limits our sample contains some 14,000
objects, i.e.\ around 3,500 in each 2dF area. Thus each region
requires a total of ten 2dF set ups (to allow for `sky' and broken 
fibres).

\begin{figure}
\psfig{file=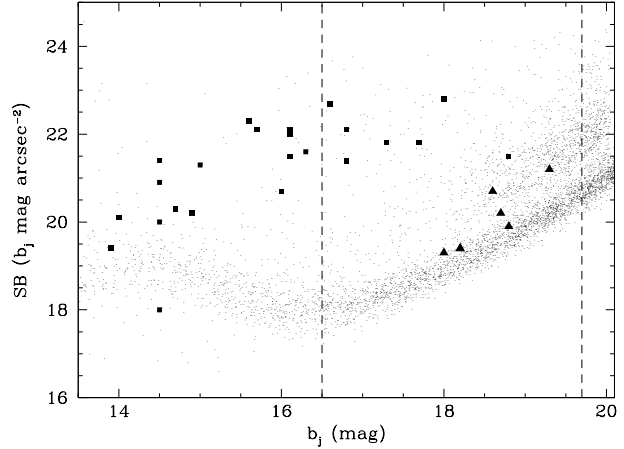,width=\one_wide}
\centering
\caption{ 
Plot of central surface brightness (SB) against \bj\ magnitude for all
objects in our APM input catalogue for the central 2dF field (points).
The unresolved objects form a locus at the lower right; 
the surface brightness (which is optimised for galaxies and faint images) 
is very poor for the unresolved
sources at bright magnitudes. 
 Objects with published redshifts are indicated by solid
symbols: galaxies (squares) and quasars (triangles). The two vertical dashed
lines indicate the magnitude limits of our survey. The selection limit for 
the input APM catalogue crosses the plotted region only in the very top
right hand corner at $\Bj \simeq 20$, SB $ \simeq 25\Bu$. 
\label{fig01_sbm}
}
\end{figure}

The selection of our targets is illustrated in Fig.~\ref{fig01_sbm}, a
magnitude-surface brightness (SB) diagram of all objects in our
central 2dF field (`Field 1') from the APM data. The APM points
include stars which follow a well-defined locus at the bottom right of
the distribution. 
The selection limit on the input APM catalogue (an area of 16 pixels at the
isophotal detection threshold) does not impinge on the area from which our
spectroscopic targets are chosen, running above the area
shown in figure 2, at SB $> 25\Bu$, except for a tiny intersection
with the top right hand corner of the plotted region at $\Bj \simeq 20$
(i.e. our objects are well within
the completeness limit of the overall APM catalogue).
We also show on the diagram the positions of the
objects discussed above with published redshifts in this same field
(very bright cluster galaxies with $\bj<13$ were not matched). Apart
from a few QSOs, the previously observed galaxies occupy a very small
part of the diagram, tending to bright magnitudes and `normal' surface
brightness. Our new survey sample defined by the dashed lines includes
the full range of surface brightness detected in the Schmidt data 
in that magnitude range. The breakdown of the sample by
image classification is given in Table~\ref{tab_nos}; as expected it
is dominated by unresolved objects (stars).

\begin{table}
\caption{Target Numbers in Field 1 with $16.5\leq\bj\leq19.7$}
\label{tab_nos}
\begin{tabular}{llll}
\hline 
    unresolved &  resolved &  merged   &  total\\
\hline 
    2152  &    785  &   307  &   3244 \\
\hline 
\end{tabular}
\end{table}

Our final spectroscopic sample will inevitably suffer from
incompleteness at the low surface brightness limit. It will not be
possible to measure spectra for the faintest LSB galaxies in
reasonable exposure times even though the multiplex advantage of 2dF
will enable us to go much fainter than previous work.

\subsection{Photometric Calibration}
\label{sec_phot}

The photometric calibration of the input catalogues is complicated by
the non-linear response of the photographic emulsion. The methods we
use to estimate the true fluxes are described in this Section. We use
different methods to measure the stars, which are often heavily
saturated, to the galaxies which are mostly unsaturated. The choice of
estimator is based on the automated APM classifications of the objects
as `stellar' or `resolved', although we emphasise again that objects
of all morphological types are observed. In all of the discussion
below we use the photographic blue \bj\ magnitude defined by the IIIaJ
emulsion combined with a GG 395 filter. This is related to the
standard Cousins $B$ magnitude by $\bj=B-0.28\times(B-V)$ (Blair \&
Gilmore 1982).

\subsubsection{Unresolved objects (`stars')}


It is relatively easy to estimate the magnitudes of unresolved objects
(probably stars) from photographic data because the images all have
the same shape. This means that the total magnitudes can be reliably
derived from the outer, unsaturated regions of the brighter
objects. The object magnitudes in the APM catalogue data (Irwin et
al.\ 1994) were measured this way using an internal self-calibration
procedure to fit stellar profiles, correcting for the non-linear
response of the photographic emulsion (see Bunclark \& Irwin 1984).
The same method was used for objects classified as `merged', since, as
discussed above, these are dominated by stars.

\begin{figure}
\psfig{file=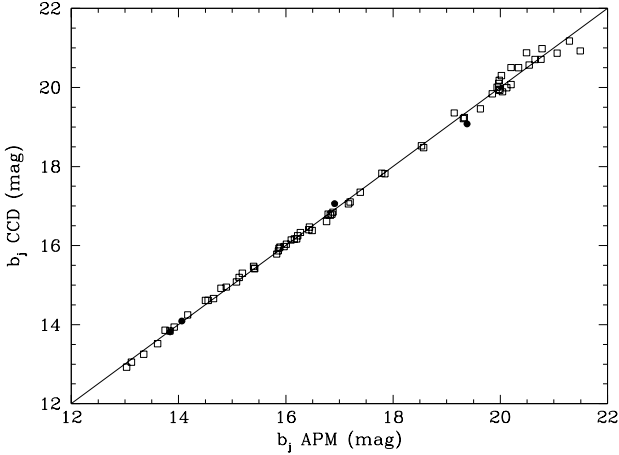,width=\one_wide}
\centering
\caption{
Photometric \bj\ calibration of unresolved objects: calibrated CCD
magnitudes are plotted against our recalibrated APM magnitudes for
Guide Star objects (solid circles) and our new CCD observations (open
squares). The line represents the identity relation between the two values.
\label{fig_cals}
}
\end{figure}

The default APM calibration uses a quadratic relation to convert from
raw instrumental magnitudes to calibrated \bj\ magnitudes. We checked
this calibration with CCD data and made a small adjustment to the
default values. We based our recalibration on 2 stars from the Guide
Star Photometric Catalog (Lasker et al.\ 1988), 2 fainter stars from
CCD frames taken for the same project (Lasker, private communication,
1997) and 68 stars from our own CCD observations with the
CTIO\footnote{Cerro Tololo Interamerican Observatory (CTIO) is
operated by the Association of Universities for Research in Astronomy
Inc. (AURA), under a cooperative agreement with the National Science
Foundation as part of the National Optical Astronomy Observatories.}
Curtis Schmidt Telescope. In each case the photographic \bj\ magnitudes 
were derived from the BV calibration data. Our adjustment to the
default APM calibration was equivalent to a shift of about 0.2 mag in
the sense that our recalibrated values are fainter. We plot the
calibrated CCD magnitudes against our final adopted magnitudes in
Fig.~\ref{fig_cals}. Our fit is very good with an r.m.s.\ scatter of
0.13 mag over a range of 8 magnitudes (see Table~\ref{tab_cal}).

\begin{table}
\caption{Photometric Calibration}
\label{tab_cal}
\begin{tabular}{llrrrr}
\hline 
Source & Type & N & mean$\Delta$ & $\sigma_\Delta$ & median$\Delta$\\
\hline 
CCD     & stars   &   72 &   0.00 &  0.13 &    0.00 \\
Caldwell& galaxies&   23 &   0.26 &  0.30 &    0.30 \\
CCD     & galaxies&  502 &$-$0.05 &  0.47 & $-$0.14 \\
FCC     & galaxies& 2033 &   0.24 &  0.52 &    0.20 \\
\hline 
\end{tabular}

\mbox{ } \\
Note: $\Delta = \bj-\bj_{calibration}$
\end{table}

\subsubsection{Resolved objects (`galaxies')}
\label{ssec_resolvedobjs}

For resolved objects (probably galaxies) we did not use the APM
catalogue magnitudes, but instead used total magnitudes estimated by
fitting exponential disk profiles to the APM image parameters (as in
Davies et al.\ 1988 and Irwin et al.\ 1990). In this section we
describe the calibration of these total galaxy magnitudes.

The absolute calibration was taken from the original calibration of
the Fornax Schmidt survey plate by Cawson et al.\ (1987). They used
CCD images of 18 cluster galaxies which they compared pixel by pixel
with the APM machine scan data of the same galaxies. The CCD images
were calibrated using standard stars and should correspond closely to
the standard Johnson B band. Cawson et al.\ quote a calibration error
of 0.1 mag. 

The Cawson et al.\ calibration was subsequently used by Davies et
al.\ (1988) for their sample of Fornax LSBGs
and for the photometry of brighter galaxies by Disney et
al.\ (1990). Ferguson (1989) carried out an independent calibration of
galaxies in the Fornax region and where his sample overlaps with that
of Davies et al.\ the mean difference between the two magnitude scales
is 0.09 mag. The APM images data for objects classified as `galaxies'
were calibrated directly from the Davies et al.\ (1988) sample (see
Morshidi-Esslinger et al.\ 1999).

At high surface brightness levels (brighter than 22.7\Bu; Cawson et
al.\ 1987; Davies et al.\ 1988) the limited dynamic range of the APM
machine affects the calculated APM magnitude, even for galaxies.  To
alleviate this problem, for each galaxy we have fitted an exponential
to the surface brightness profile in the range between $\mu_{B} =
22.7\Bu$ and the limiting surface brightness (detection isophote)
$\mu_{L} = 25.7\Bu$.  This procedure largely overcomes the problems of
saturation at the centre of an image, but of course will not allow for
any central excess light such as a nucleus. We chose to use an
exponential to fit each surface brightness profile because a large
fraction of the galaxy population is well fitted by such a function
(Davies et al.\ 1988). The exponential fit gives values for the
extrapolated central surface brightness $\mu_{x}$ and the exponential
scale length $\alpha$. From these the relation $m_{tot} = \mu_{x} -
5\log(\alpha) - 1.995$ can be used to derive the total apparent
magnitude $m_{tot}$ under the fitted profile.  The surface brightness
profile data are supplied as part of the APM images file. (In fact,
the image area at different surface brightnesses is supplied, and this
is used to produce a surface brightness profile assuming circular
isophotes; see Phillipps et al.\ 1987, Morshidi-Esslinger et al.\
1999).

\begin{figure}
\psfig{file=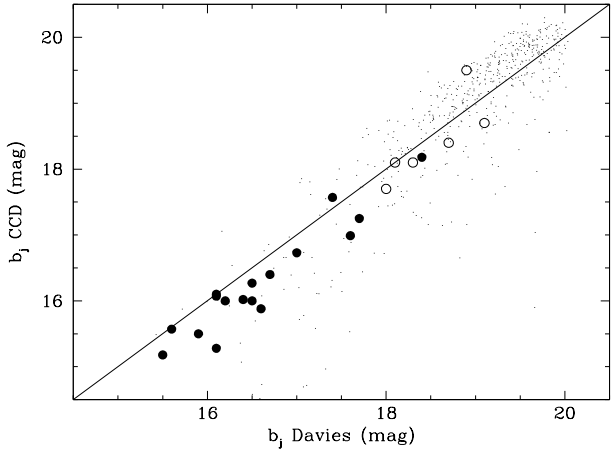,width=\one_wide}
\centering
\caption{ Photometric \bj\  calibration of resolved objects: calibrated
CCD magnitudes are plotted against our fitted photographic magnitudes
(Davies et al.\  1988) for
galaxies measured by Caldwell \& Bothun (1987) (solid circles; open
circles indicate galaxies for which we assumed $B-V=0.72$) and our new
CCD observations (points). The line represents the identity relation
between the two values.
\label{fig_calg}}
\end{figure}

We compare our galaxy photometry with the CCD photometry of Caldwell
\& Bothun (1987) in Fig.~\ref{fig_calg}. Here we have plotted total
\bj\  magnitudes from Caldwell \& Bothun against our photographic
galaxy magnitudes. Following Ferguson \& Sandage (1988) we have
assumed $B-V=0.72$ when converting to \bj\ if the colour was not
measured by Caldwell \& Bothun. Over the whole range shown there is an
average offset of about 0.2 mag in the sense that our magnitudes are
fainter. We found a similar offset when comparing our whole sample to
that of Ferguson (1989), also measured from photographic plates (not
plotted, but listed in Table~\ref{tab_cal}).  We also estimated total
(Kron) galaxy magnitudes using our own CCD data for a larger sample of
faint galaxies: these show a small average offset in the opposite
sense (see Table~\ref{tab_cal}). When plotted in Fig.~\ref{fig_calg}
these data give the impression that the slope of the calibration is
steeper than unity, but this is also consistent with a constant offset
in the Caldwell \& Bothun data points as they are all at brighter magnitudes.
We have retained the Davies et al.\ (1988) calibration without any
further adjustment since we do not expect a closer agreement from
methods which use different data and fit different profiles to
estimate total galaxy magnitudes.

\subsection{Astrometric Calibration and Guide Stars}


The input target lists (including guide stars) must have relative
positions accurate to 0.3\asec\ or better over the full two degree
field of the 2dF spectrograph as discussed above. This condition is
not satisfied by all image catalogues based on UKST plates---see
Drinkwater, Barnes \& Ellison (1995) for a discussion of the
problems. We therefore checked the accuracy of the APM positions by
comparison with the Positions and Proper Motions star catalogue (PPM;
R\"{o}ser, Bastian \& Kuzmin, 1994).

A total of 232 PPM stars matched stars in the APM catalogue file for
the whole UK Schmidt field F358 with \bj\ magnitudes of 8.9--12.1. We
eliminated outliers (total position errors more than 1.5 arcsec) and
selected a test sample of the faintest 60 remaining stars. The mean
and r.m.s. errors were $(-0.39\pm0.30)$ arcsec in RA and
$(-0.12\pm0.23)$ arcsec in Dec. These errors are for the whole 6 degree
field of the UK Schmidt: in a single 2dF field the scatter is about
0.25 arcsec in each direction.  We also calculated the radial errors
as a function of radius from the plate centre as in Drinkwater et al.\
(1995) to test for any overall scale errors. We found a very small
error: the faint stars were slightly too far from the centre (0.089
arcsec/degree) and the bright stars were slightly too close to the
centre ($-0.021$ arcsec/degree). This is in the same sense as we found
for the COSMOS/UKST catalogue, but smaller in magnitude: there we
measured 0.14 and $-0.29$ arcsec/degree respectively. This shift (the
`magnitude' effect) between the bright and faint stars is caused by
asymmetric halos around the brighter stars which displace their image
centroids away from the plate centre. The largest
scale error we found would not be significant over a two-degree field,
so we have established that APM catalogue positions are accurate
enough for 2dF observing.

The guide stars used to align the telescope during 2dF observations
must have positions to the same accuracy as the targets
and---importantly---in the same reference frame. We chose stars from
the same APM catalogue file as our targets but took care to minimise
errors that could arise from proper motion or the `magnitude
effect'. Both these problems are reduced by selecting faint guide
stars, but we can further reduce the chance of selecting stars with
significant proper motions by choosing stars with blue colours and APM
positions measured from the more recent red survey plates (see
Drinkwater, Barot \& Irwin 1996). We used faint blue stars with
$15.5\leq\bj\leq16.5$ and $B-R<1.8$. As a rule-of-thumb, this corresponds to
stars with barely discernible diffraction spikes on the blue survey
plates and no diffraction spikes on the red survey plates. We found
that stars selected according to these criteria were easily detected
by the 2dF guiding system and had consistent positions.

\section{Scientific Rationale} 
\label{sec_science}

As emphasised in the Introduction, the \FSS\ will cover all
possible types of objects visible within our target magnitude
range. As such, it will be useful for a large number of individual
survey projects. In this Section we summarise some of these, 
divided according to object type.

\subsection{Cluster Galaxies}
\label{sec_cluster}


The prime reason for choosing the sky area we have was, of course, the
presence of the Fornax Cluster of galaxies. The Fornax and Virgo 
galaxy clusters are the nearest reasonably rich clusters (Fornax is 
approximately Abell Richness Class~0~-- it is supplementary 
cluster S0373 in Abell, Corwin \& Olowin 1989). 
We take the distance to Fornax to be 15.4 \Mpc\ (a
distance modulus of 30.9\mag) as derived by Bureau, Mould \&
Staveley-Smith (1996), so our 2dF sample reaches absolute
magnitudes around $M_{B} = -11$. The Fornax Cluster 
has been the subject of several previous
spectroscopic (Jones \& Jones 1980; Drinkwater \& Gregg 1998;
Hilker et al. 1999) and
photometric studies (Phillipps et al.\ 1987; Davies et al.\ 1988;
Ferguson 1989; Ferguson \& Sandage 1988; Irwin et al.\ 1990;
Morshidi-Eslinger et al. 1999).

The main motivation for an all-object survey is, of course, to
determine cluster membership for a complete sample of objects (including,
especially, dwarf galaxies),
irrespective of morphology. This will allow us to 
test whether the usual assignment of LSBGs to the
cluster and high (or even `normal') surface brightness faint
galaxies to the background is justified 
(see, for example, the contrasting views expressed in Ferguson \& Sandage
1988, and Irwin et al.\ 1990)  
and enable us to determine the complete surface brightness
distribution (and the joint surface brightness - magnitude distribution)
for a cluster for the first time (see Phillipps et al.\
1987; Phillipps 1997). We report elsewhere on the identification of a
population of low luminosity, very compact HSBGs in the cluster
(Drinkwater et al.\ 2000b = Paper IV).

\subsection{Normal Field Galaxies}
\label{sec_field}


Our galaxy sample will, though, be dominated by very large numbers
of background galaxies. These will be roughly 10 times more numerous
than the cluster members (depending on the relative slopes of the
field number counts and the faint end of the cluster luminosity function) and
can obviously be used to determine a field
luminosity function (LF), using the conventional approach which requires
redshift data. Although containing many fewer galaxies than the major
2dF or Sloan surveys, an LF determined from the \FSS\ sample will
have the advantage of including all galaxies, irrespective of
morphology, at both the high and low surface brightness ends (down,
obviously, to the surface brightness limit of the 2dF observations).
Surface brightness limitations on the determination of LFs have been
discussed by Phillipps \& Driver (1995) and Dalcanton (1998). 
Extending this, we will be able to determine the bivariate brightness
distribution for field galaxies
(i.e. the joint distribution in luminosity and surface brightness;
Phillipps \& Disney 1986, van der Kruit 1987, Boyce \& Phillipps 1995,
de Jong 1996). 

\subsection{Compact and LSB Field Galaxies}
\label{sec_celg}


We have already detected a number of very compact field galaxies
beyond the Fornax Cluster (see Paper~II for details). These
objects are so compact that they have been classified as `stars' from
the blue sky survey plates and therefore represent a class of galaxy
missed in previous galaxy surveys based on photographic survey
plates. 
We estimate that they represent $2.8\pm1.6\%$ of all galaxies in the
magnitude range $16.5\leq\Bj\leq19.7$.
They are luminous (within a few magnitudes of $M_*$) and
most have strong emission lines and small sizes typical of luminous
\hii\ galaxies and compact narrow emission line galaxies. 
Four of the thirteen have red
colours and early-type spectra, so are of a type
unlikely to have been detected in any previous surveys.

Similarly we have been able to obtain spectra for a number of LSBGs
beyond Fornax. Some are likely to have been misclassified as cluster members
on morphological grounds (Ferguson 1989) and can only be revealed as
larger background LSBGs (potential `Malin 1 cousins'; Freeman 1999)
via redshift measurements
(Drinkwater et al.\ 1999b; Jones et al.\ 2000 = Paper~III, in preparation).

\subsection{Galactic Stars}


Although initially motivated by extragalactic interests, the \FSS\
can also make significant contributions to Galactic astronomy.  The
lion's share of the unresolved targets in the survey will be ordinary
Galactic stars, making up  
about 70\% of the objects in the overall survey.
For instance, the final tabulation will include many thousand M dwarfs in the
Galactic disk.
While the \FSS\ 2dF velocity
precision ($\sim 50\kms$) is low compared to that used in most
kinematic studies of the Galaxy (e.g.\ Norris, 1994), the sheer
numbers of M dwarfs should allow a good determination of their scale
height and velocity dispersion, for example.
As Fornax is only $\sim 30\degr$ from the South Galactic Pole, many
of the stars in the survey will belong to the halo.  Although only a
minor mass component of the Galaxy, the properties of the halo provide
clues to the formation of the whole Milky Way.  Blue horizontal branch
stars from the metal poor halo will make up perhaps $\sim 1\%$ of our
sample and are straightforward to recognise spectroscopically.  

\subsection{Radio Sources}

The region of our \fss\ is covered by two sensitive radio continuum
surveys -- the NRAO VLA Sky Survey (NVSS; Condon et al.\ 1998) and the
Sydney University Molonglo Sky Survey (SUMSS; Bock, Large and Sadler
1999).  These cover different frequencies (1.4\,GHz for NVSS, 843\,MHz
for SUMSS), both with an angular resolution of about 45\asec.
The faintest radio sources catalogued by these surveys are roughly
2.5\,mJy for NVSS and 5\,mJy for SUMSS. At these faint flux density
levels we expect to detect three main kinds of radio sources: QSOs,
active galaxies and star-forming galaxies (Kron, Koo \& Windhorst
1985, Condon 1984; see also Condon 1992). The fraction of
star-forming galaxies increases rapidly below 10\,mJy, and below
1\,mJy they become the dominant radio--source population Our 2dF spectra
should discriminate reliably between AGN and starburst galaxies.

\subsection{Quasi-Stellar Objects}

As well as the foregoing radio quasars, the survey will detect
one of the largest ever completely unbiased samples of optical QSOs. 
All previous
optical QSO surveys have relied on one or more specific selection
criteria, such as UV-excess or variability, to pre-select `candidate' QSOs 
for follow-up spectroscopy (e.g. Boyle, Jones \& Shanks 1991).
The {\it FSS} on the other hand will be limited only by the strength of
the QSO's emission lines. Preliminary results suggest that this
technique detects some 10\% more QSOs to the same magnitude limit as
conventional multi-colour work (see Meyer et al.\ 2000 = Paper~V).
\section{Spectroscopic Observations}
\label{sec_obs}


In this Section we describe our spectroscopic data from the 2dF. We
give a brief summary of the observing setup and initial data reduction
and then explain the semi-automated analysis we perform to classify all the
spectra. Finally we present some example spectra from our initial
observations.

\subsection{Observing Setup}

We observed all our targets with the same observing setup for 2dF: the
300B grating and a central wavelength setting of 5800\AA\ giving a
wavelength coverage of 3600--8010\AA\ at a resolution of 9\AA\ (a
dispersion of 4.3\AA\ per pixel). This is the same setup as for the
2dF galaxy (Folkes et al. 1999) and QSO (Boyle et al. 1997)
redshift surveys. We did not attempt to flux-calibrate our spectra 
given the difficulty of flux-calibration in fibre-fed spectroscopy, and 
because our objective was to measure velocities for as many objects 
as possible in the available time. 

In order to maximise our observing efficiency we grouped our targets
by their central surface brightness, so that we could vary exposure 
times to obtain similar quality spectra over a large range of 
apparent surface brightness. The exposure times ranged from 
30 minutes for bright stars to four hours for
LSB galaxies.  Our early runs in commissioning time were limited to
minimum exposures of about 2 hours and also included a range of
objects and surface brightnesses, so produced some very high 
signal-to-noise stellar
spectra. We discuss the quality of the spectra as a function of
exposure time and surface brightness in Section~\ref{sec_results}.

\subsection{Data Reduction}

The 2dF facility includes its own data reduction package ({\small 2DFDR})
which permits fast, semi-automatic reduction of data direct from the
instrument. When we started the project {\small 2DFDR} was still under
development, so we chose instead to reduce the data with the {\small
DOFIBERS} package in IRAF\footnote{IRAF is distributed by the National
Optical Astronomy Observatories, operated by the Association of
Universities for Research in Astronomy, Inc.  under cooperative
agreement with the NSF.}. We are now reducing the data in parallel
with {\small 2DFDR} to compare our results.

The data reduction with IRAF follows the standard procedures for
multi-object fibre spectroscopy supplemented by several scripts used
to reformat the image header data and tabulate the object
identifications using the output from the 2dF configuration software and
the 2dF fibre-spectrum lookup table.

Accurate sky-subtraction with fibre spectra is difficult and problematic
(Barden et al. 1993; Watson, Offer \& Lewis 1998) especially for the
stronger night sky-lines and when the fibres 
are closely spaced with profiles only 2-3 pixels wide (as with 2dF).
After extracting the spectra we removed residuals from the strong
night sky lines at 5577 and 6300\AA\ by interpolation across the
lines.  The spectra were then visually inspected and any strong
features due to cosmic ray events removed: these were identified by
having widths less than the instrumental resolution.

We also remove atmospheric absorption features from the spectra using
a simple self-calibration method. We take all the galaxies observed
with a given CCD on a given night and average them with no
weighting. The galaxy features cancel out as they are all at different
redshifts, leaving a combination of the instrumental response and the
main atmospheric absorption bands. We then fit and normalise by a
continuum. Then we edit the resulting spectrum by hand to set it to
unity in all regions except the main atmospheric bands at
6800--7600\AA. We then divide all object spectra by this to remove the
bands. This has the effect of removing the atmospheric features, but
otherwise leaving the spectra unchanged with instrumental response
intact. The same approach could not be used when most of the spectra
on a given CCD are stars with many common features all at the same
wavelengths. For these we found that a normalisation spectrum
generated from galaxies observed the same night was more than
adequate.

\subsection{Spectral Analysis}
\label{ssec_spectra}

The aim of our spectral analysis is to determine a redshift and
identification for all spectra ranging from Galactic stars to high
redshift QSOs. In keeping with our survey philosophy we analyse all
the spectra in an identical fashion, irrespective of their image
morphology. To do this successfully we have adapted the usual
procedure in galaxy surveys of cross-correlating against template
galaxy spectra by using a set of stellar templates instead of the
normal templates of absorption-line galaxies. This was previously used
on a small sample of unresolved objects by Colless et al.\ (1993).

The first stage of the identification is to calculate
cross-correlations automatically using the IRAF add-on package {\small
RVSAO} (Kurtz \& Mink, 1998).  For each spectrum-template combination
this measures the Tonry \& Davis (1979) $R$ coefficient, the redshift
and its error. Emission lines are not removed before performing the
correlations, but the spectra are Fourier-filtered. At this stage the
full available wavelength range is used for the cross-correlations.
The redshifts are measured as radial velocities in units of $cz$ and
are subsequently converted to heliocentric values.  By chosing the
template giving the best $R$ coefficient we can determine not only the
redshift, but a first estimate of the object type. We only accept
identifications with $R\ge3$ (corresponding, in principle, to a peak in the
cross-correlation which is significant at the 3 sigma level) and in
addition make a check by eye for misidentifications (see below). This is 
straightforward for $R \ge 3$, since in practice such spectra always have
three or more identifiable features.
Objects with redshifts of $\approx500\kms$ or less are
Galactic stars for which the best template indicates the stellar
spectral type.  At higher redshifts, external galaxies are separated
into absorption-line types if they match one of the stellar spectra or
emission-line types if they match the emission-line galaxy template
best. We found that all the absorption-line galaxies that would have
been detected by galaxy templates were easily measured using the
stellar templates, so we do not need to use specific absorption-line
galaxy templates.

\begin{figure}
\psfig{file=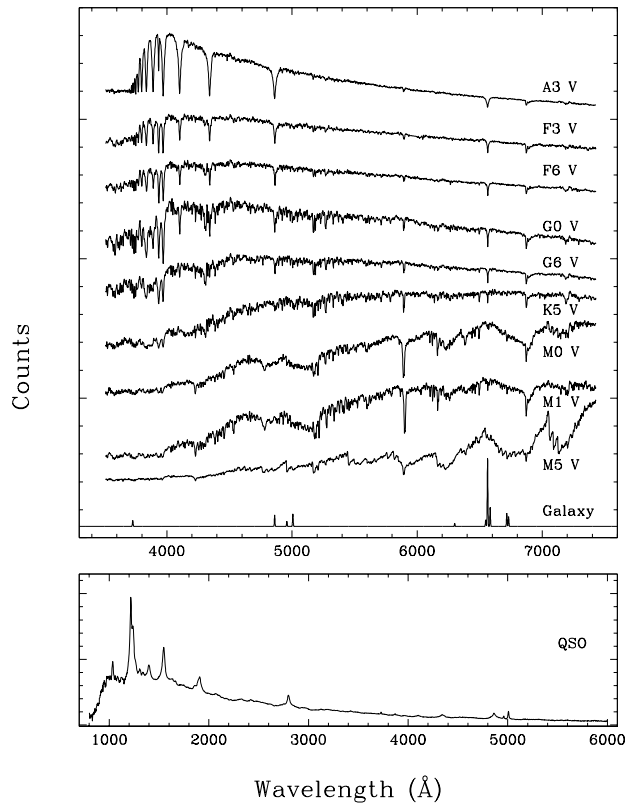,width=\one_wide}
\centering
\caption{
Spectra of the various templates used for automatic classification
and redshift estimation. See Table~\ref{tab_temp} for details.
\label{fig04_tem}
}
\end{figure}

The template spectra used are listed in Table~\ref{tab_temp} and plotted 
in Fig.~\ref{fig04_tem}: we use a set of nine stellar templates from the 
Jacoby, Hunter \& Christian (1984) library and a synthetic emission-line 
galaxy spectrum provided with the {\small RVSAO} package. We constructed 
a second emission-line template similar to the first but limited to 
wavelengths less than 6000\AA. This was needed to
give reasonable fits to the high-redshift galaxies where the \ha\ line
was shifted out of the 2dF bandpass, but strong \hb/\oiii\ features
were present. The stellar templates were chosen to give a reasonable
range of spectral types, but not more than could be separated with our
low-resolution unfluxed spectra. We also note that the Jacoby et al.\
spectra have not been shifted to zero redshift: we therefore estimated
their redshifts using a combination of individual line measurements
and cross-correlation with other standards. These were then entered in
the image headers to give the correct results with {\small RVSAO};
they are also listed in Table~\ref{tab_temp}.

\begin{table}
\caption{Templates\label{tab_temp}}
\begin{tabular}{llr}
  \hline 
  Name          & Type & Velocity\\
                &      & (\kms)\\
  \hline 
  HD 221741     & A3 V &  75  \\
  Hz 948        & F3 V & 105  \\
  SAO 57199     & F6 V & 170  \\
  HD 28099      & G0 V & 110  \\
  HD 22193      & G6 V & 140  \\
  SAO 76803     & K5 V &  95  \\
  HD 260655     & M0 V &  -8  \\
  BD 63 0137    & M1 V & 410  \\
  Yale 1755     & M5 V &  57  \\
  Emission Line & galaxy  &   0  \\
  LBQS QSO      & QSO  &   0  \\
  \hline 
\end{tabular}
\end{table}

In the second stage of the identification process we check each
identification interactively using the {\small RVSAO} package to
display the best cross-correlation and a plot of the object spectrum
with common spectral features plotted at the corresponding redshift.
When the redshift is obviously wrong (e.g.\ with the Calcium H \& K
lines clearly present but misidentified), it is flagged as being wrong
or in some cases is recalculated. The recaculation most commonly
involves repeating the template cross-correlation on a restricted
wavelength range chosen to avoid the red end of the spectrum affected
by poorly removed sky features.  In extreme cases the object may be a
QSO: these are distinguished by strong broad emission lines and are
measured using a composite QSO spectrum (Francis et al.\
1991). Objects still not identified at this stage are flagged to be
reobserved.

A third, supplementary stage is used for any spectra measured with
good signal (a signal-to-noise ratio $>$10 in each 4.3\AA\ wide pixel)
but no obvious features in their 2dF spectra: these are flagged as
`strange' and scheduled for detailed follow-up observations with
conventional slit spectrographs.

Once the spectroscopic redshift measurements are complete, they are
corrected to heliocentric values. We checked the accuracy of the
redshift measurements by comparing the results for 66 objects with
repeated measurements. The r.m.s.\ scatter of the velocity 
differences is 90\kms.
This uncertainty is consistent with the combined error estimates for
the same measurements produced by {\small RVSAO}: the mean predicted
error was 92\kms. Note that this implies a measurement error of a
single observation of $90/\sqrt{2} \simeq 64 \kms$. We also compared
our results to redshifts of 44 galaxies found in a search of the
literature using NED (most were from Hilker et al.\ 1999). The
comparison gave a mean velocity difference of ($7\pm17$)\kms\ and an
r.m.s.\ scatter of 111\kms, entirely consistent with our internal
calibration.

The 2dF spectra, although of low resolution and unfluxed, are useful
for more detailed analysis than simple redshift measurements and
object classifications (c.f. Tresse et al. 1999).  
We defer any detailed analysis of the spectra
to later papers dealing with specific object classes, but note here
that, even for the lowest luminosity galaxies,
they can be used to measure emission line equivalent widths, 
and hence star formation rates, line
widths (limited by the resolution of 900\kms), emission line ratios
(e.g.\ \oiii/\hb\ and \nii/\ha), absorption line indices (e.g.\ \caii
H+\he/\caii K and \hd/Fe{\small I}4045) and even ages and metallicities
from these Balmer and the metal absorption lines
(Paper II; Folkes et al. 1999).

\subsection{Current status}
\label{sec_results}

Although this present paper is mainly concerned with the principles
behind the \FSS, we already have a considerable amount of 2dF data for
the project from commissioning observations in 1996/1997, and
scheduled time in December 1997/January 1998 and November 1998. We
have nearly completed our observations of the first field having
observed 92\% of all targets in the range $\bj = 16.5$ to 19.7 and
successfully obtained redshifts for 94\% of those observed.

\begin{figure}
\psfig{file=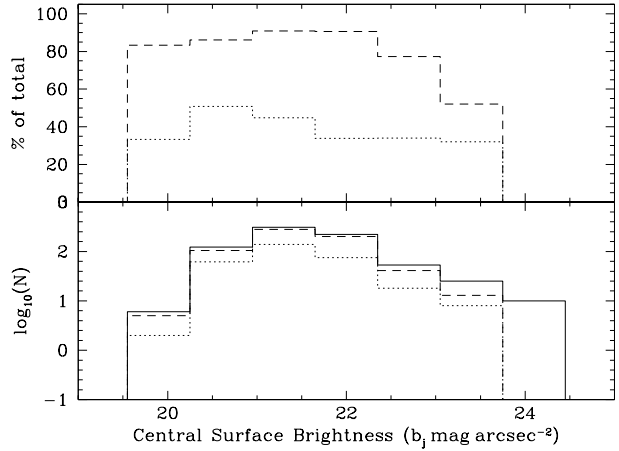,width=\one_wide}
\centering
\caption{
The number of galaxies (resolved objects) observed which have successful
redshift measurements as a function of surface brightness. The lower panel
shows the total number of galaxies observed (solid line), the number
with measured redshifts (dashed line) and the number of 
those which have strong emission lines (dotted). The upper panel
again shows the successfully measured (dashed) and emission line (dotted)
galaxies, this time as a fraction of those observed.
\label{fig05_lim}
}
\end{figure}

For resolved objects (galaxies) the success rate of our redshift
measurements is a function of surface brightness.
In Fig.~\ref{fig05_lim} we plot the numbers of galaxies observed and
identified as a function of central surface brightness. We have
attempted to optimise the exposure times to the surface brightnesses
of the objects, using exposures up to 3.75~hours for the lower surface
brightness images. The identification rate runs at 78\% or better to a
limit of 23\Bu.  Fainter than this limit (corresponding to a mean
surface brightness inside the detection threshold $\sim 24.5\Bu$), the
identifications drop off rapidly. The unresolved objects at higher
surface brightness (mostly stars) have an identification rate of 95\% in our
target magnitude range of $\bj = 16.5$ to 19.7.

\begin{figure}
\psfig{file=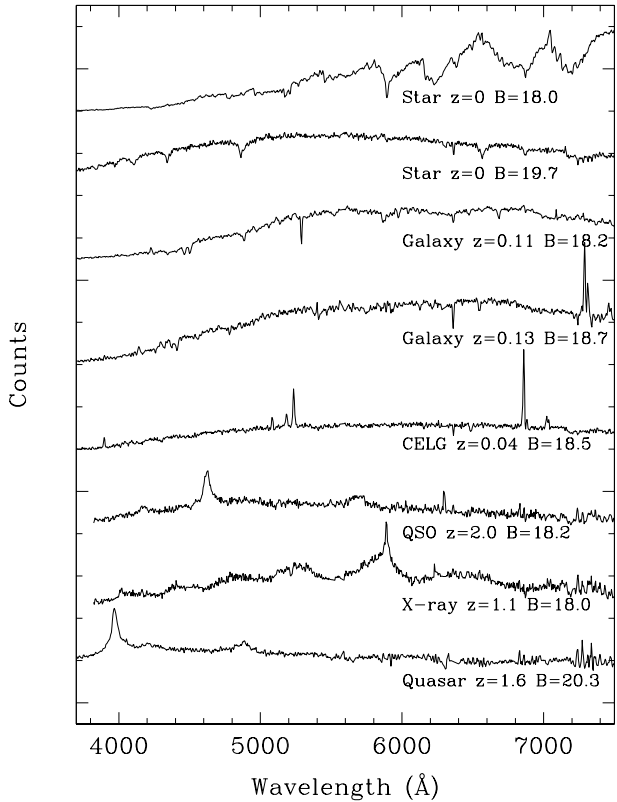,width=\one_wide}
\centering
\caption{ Example spectra of stars, galaxies and quasars from our
initial 2dF observations. See text for a brief description of the
spectra.
\label{fig06_spc}
}
\end{figure}

In Fig.~\ref{fig06_spc} we show example spectra from our initial
observations of the various types of object discussed above.  The
first two spectra are Galactic stars, an M-dwarf and a white
dwarf. The next two spectra are of normal low-redshift galaxies, one
with an absorption line spectrum and one with an emission line
spectrum. The remaining four spectra are all of objects that were
unresolved (i.e.\ classified as stars in the target catalogues), but
have been identified as various types of galaxy. The first is a
compact emission line galaxy (CELG; see Section~\ref{sec_celg}), the
second is a normal, optically selected QSO and the third is an X-ray
source. The final spectrum is of a fainter radio-loud quasar.

\section{Initial Scientific Results: velocity distributions from the 
 first 2dF field}
\label{sec_initsci}

The number of galaxies observed in the Fornax Cluster itself 
is not yet large enough to allow a detailed study of the cluster 
population: this will require results from the remaining three 
2dF spectrograph fields to achieve the statistical samples needed. 
However, we have ample data 
to delineate clearly the velocity structure in the direction
of Fornax. In particular, we have determined accurately the velocity 
distribution of Galactic stars, as well as the
galaxy distribution in redshift space behind the cluster.

\subsection{Radial velocities of Galactic stars} 
\label{sec_is_stars}

The radial velocity distribution of Galactic stars is revealed 
in the existing {\em FSS} results, despite the modest 
resolution of the spectra compared to those conventionally used 
in kinematic surveys of the Galaxy. 
A total of 2467 objects in Field~1 of Table~\ref{tab_fields} 
have reliable radial velocities $v_r < 750\;\mbox{km} \mbox{s}^{-1}$. 
The estimated standard errors in the velocities are typically 
$\pm\:25-80\:\mbox{km}\:\mbox{s}^{-1}$, 
sufficiently small to reveal the contributions of different Galactic 
components. 

\begin{figure}
\psfig{file=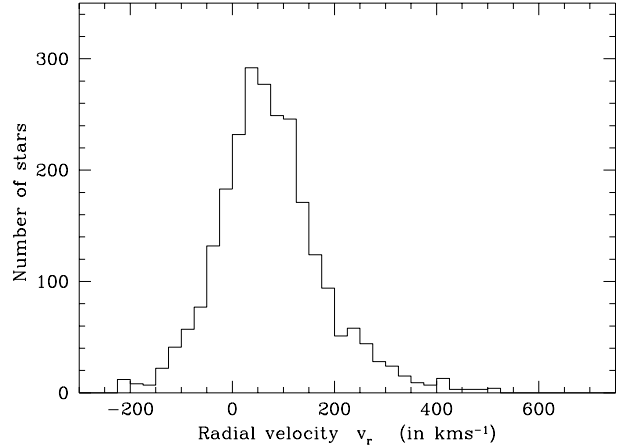,width=\one_wide}
\centering
\caption{ The heliocentric radial velocity distribution for Galactic stars.
\label{fig_is_starvels}
}
\end{figure}

Figure~\ref{fig_is_starvels} shows the distribution of all 
FSS Field~1 objects 
with reliable ($R\geq3$) velocities less than 750~kms$^{-1}$. 
The field has Galactic co-ordinates 
$(l,b) = (237^{\rm o},-54^{\rm o})$, so we are 
sampling a sight-line looking diagonally `down' through the Galactic 
plane between the anti-centre and anti-rotation directions.
The component of the motion of the local standard of rest 
in this direction is $-120$~kms$^{-1}$. 
For our chosen magnitude range, the survey will sample predominantly 
disc, thick disc and halo main sequence stars, with some 
contribution from halo giants and disc white dwarfs (Gilmore \& 
Reid 1983). The results can be compared with dedicated spectroscopic 
studies of faint stars in high-latitude fields (e.g. Kuijken \& Gilmore 
1989; Croswell et al. 1991; Majewski 1992). 

\begin{figure}
\psfig{file=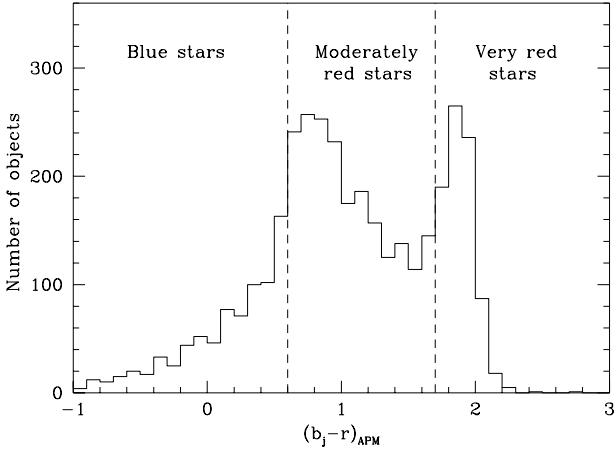,width=\one_wide}
\centering
\caption{ The colour distribution for Galactic stars. The histogram 
shows the distribution of the $(b_j-r)_{\rm APM}$ colour index 
for the stars of FSS Field~1 having reliable velocities. 
The colour limits of the subsamples of stars are indicated by 
dashed lines. 
\label{fig_is_starcols}
}
\end{figure}

   The contribution of the various Galactic components can be 
demonstrated by considering subsamples of the stars defined by 
colour. 
Basic colour information can be derived from the blue and red magnitudes
given in the APM Catalogue. Figure~\ref{fig_is_starcols} shows the 
distribution of these $(b_j-r)_{\rm APM}$ colours for the Field~1 objects with 
velocities $v_r \leq 750$~kms$^{-1}$. 
The form of the distribution is similar to that obtained in dedicated 
studies of the properties of faint stars (e.g. Reid \& Majewski 1993). 
We divide the stars into three samples: 
relatively blue stars having $(b_j-r)_{\rm APM} \leq 0.6$; 
moderately red stars having $0.6 < (b_j-r)_{\rm APM} \leq 1.7$; 
and very red stars having $(b_j-r)_{APM} > 1.7$.
The sharp decline in numbers bluewards of $(b_j-r)_{\rm APM} \simeq 0.6$ 
is the result of the main sequence cut-off for moderately old stellar 
populations; the blue sample extends to this limit. 
These limits at $(b_j-r) = 0.6$ and 1.7 correspond to 
$({\rm B}-{\rm V}) \simeq 0.4$ and 1.1 


\begin{figure}
\psfig{file=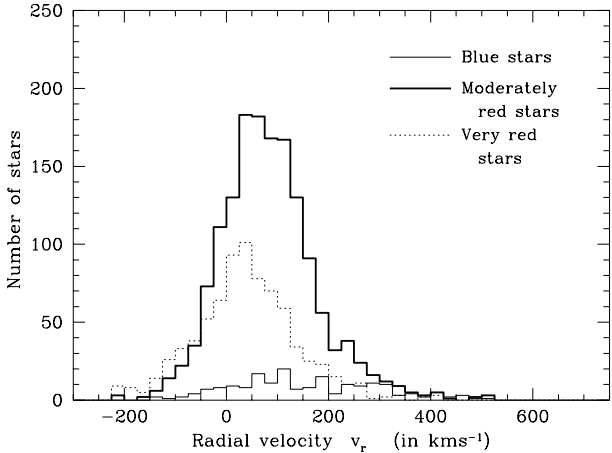,width=\one_wide}
\centering
\caption{ The radial velocity distribution for Galactic stars 
segregated by colour. 
As Fig.~\ref{fig_is_starvels} but showing the heliocentric radial velocities 
distinguished according to the APM $(b_j-r)$ colour index. 
The stars are divided into blue ($(b_j-r)_{\rm APM} \leq 0.6$), 
moderately red ($0.6 < (b_j-r)_{\rm APM} \leq 1.7$) and very 
red ($(b_j-r)_{\rm APM} > 1.7$) samples. 
\label{fig_is_starvelscols}
}
\end{figure}

The moderately red stars are expected to include G and K dwarfs 
in the thick disc and halo, G and K giants in the halo, and 
disc K dwarfs. 
The halo component, being dynamically pressure supported, has a 
broad radial velocity distribution which is displaced relative 
to the solar motion by the component of the solar rotation 
velocity towards Fornax (Freeman 1987; Gilmore, Wyse \& Kuijken 1987; 
Majewski 1993). 
Disc and thick disc stars, being rotationally supported, have a zero 
or small asymmetric drift and a modest intrinsic velocity dispersion: 
their velocity distributions will be centred closer to zero 
heliocentric velocity. The moderately-red star sample is therefore 
expected to have a broad velocity distribution with the halo component 
contributing a high velocity tail, consistent with the velocity distributions 
shown in Figure~\ref{fig_is_starvelscols}. 
In contrast, the very red star sample will be rich in disc late K and M 
dwarfs and will include halo late K and M giants. 
It is therefore expected to have only a modest net drift with respect to 
the local standard of rest but with a tail to high velocity, 
as observed in Figure~\ref{fig_is_starvelscols}. 
The blue stars include local (disc) white dwarfs and halo horizontal 
branch stars. They are therefore expected to have a broad range 
of velocities, consistent with the results here.

Of particular interest is the high velocity tail at 
$v_r \geq 400$ km s$^{-1}$. 
If the extreme examples are confirmed by higher resolution spectroscopy
they will provide useful constraints on the mass of the Galaxy
(e.g. Carney, Latham \& Laird 1988; Croswell et al. 1991; 
Majewski, Munn \& Hawley 1996; Freeman 1999, private communication).

\subsection{Galaxies in the foreground of the Fornax Cluster}
\label{sec_is_fnxfore}

A gap is present in the velocity distribution between the cut-off 
in Galactic stars at $cz \simeq 550$~kms$^{-1}$ and the Fornax 
Cluster at $cz \simeq 900 - 2200\:\mbox{km}\:\mbox{s}^{-1}$. 
No objects are found in this intermediate velocity range among the results 
from the first 2dF field. The low velocity limit of the 
cluster velocity distribution is therefore defined without 
ambiguity. 

It is of interest to determine whether there are any galaxies in the 
foreground to the Fornax Cluster having heliocentric radial velocities 
$cz < 600\:\mbox{km}\:\mbox{s}^{-1}$ which might be overlooked 
given the very large number of Galactic stars in this velocity 
range. The APM Catalogue (used as the input database for the survey) 
provides a classification for each image from the blue and red sky 
survey plates. Of the 2467 objects having 
$cz < 600\:\mbox{km}\:\mbox{s}^{-1}$ 
and cross-correlation $R$~parameter $\geq 3.0$, 
14 are classified as being galaxies in 
both blue and red. All 14 were inspected visually on the 
Digitised Sky Survey and again on a SuperCOSMOS measuring 
machine (Miller et al. 1992) scan of film OR17818 taken on 
Tech Pan emulsion with the UKST. 
The Tech Pan data provided higher resolution and greater depth than 
the Digitised Sky Survey (e.g. Phillipps \& Parker 1993). 
All foreground galaxy candidates appeared to be compact images 
merged with another, fainter image. Most were unambiguously Galactic 
stars merged with either another star or with a background galaxy. 
None of the 14 candidates had the extended appearance 
expected of a nearby dwarf galaxy. 

To extend the search, the visual inspection was repeated on the five 
images with reliable velocities 
\mbox{$\leq\:600\:\mbox{km}\:\mbox{s}^{-1}$}
having the largest APM $\sigma$ parameter on the blue sky survey plates. 
The $\sigma$ parameter measures the degree to which an image differs from a 
point-spread function and is a convenient indicator of a non-stellar 
light profile. Of the five images having a blue $\sigma_{\rm B} > 38.0$, 
none had the appearance expected of a nearby dwarf galaxy: all were 
found to be compact (star-like) images merged with either another 
star or a faint galaxy. 

A third and final search for foreground galaxies was performed using large 
exponential scale lengths as a indicator of extended images. 
The surface photometry described in Section~\ref{ssec_resolvedobjs} 
derived a scale length from the low surface brightness regions 
of each image. Only five images with reliable velocities 
$\leq\:600\:\mbox{km}\:\mbox{s}^{-1}$ had scale lengths 
$\alpha\:>\:1.5\:\mbox{arcsec}$. None had the appearance expected of 
a nearby dwarf galaxy on the Digitised Sky Survey or the scan 
of the Tech Pan film: all objects were again found to be merged images. 
We conclude that no foreground galaxies were found with star-like 
velocities.

We therefore have no galaxies with heliocentric 
\mbox{$cz\:<\:900\:\mbox{km}\:\mbox{s}^{-1}$} 
in Field~1 of Table~\ref{tab_fields} within our magnitude range. 
Among the brighter galaxies ($B < 16$) in the whole cluster region,
Jones \& Jones (1980) previously found a small number 
with such low velocities (NGC~1375, NGC~1386, NGC~1396 ($\equiv$~G75), 
and NGC~1437A), though the exact number depends on the accuracy 
of their redshift determinations. A search of the 
{\em NASA Extragalactic Database} ({\em NED}) identifies the same 
four galaxies. Of these, NGC~1375, NGC~1386 and NGC~1396 lie in 
our Field~1.

\subsection{The Velocity Structure of the Fornax Cluster}
\label{sec_is_fnxcl}

\begin{figure}
\vspace*{-10mm}
\psfig{file=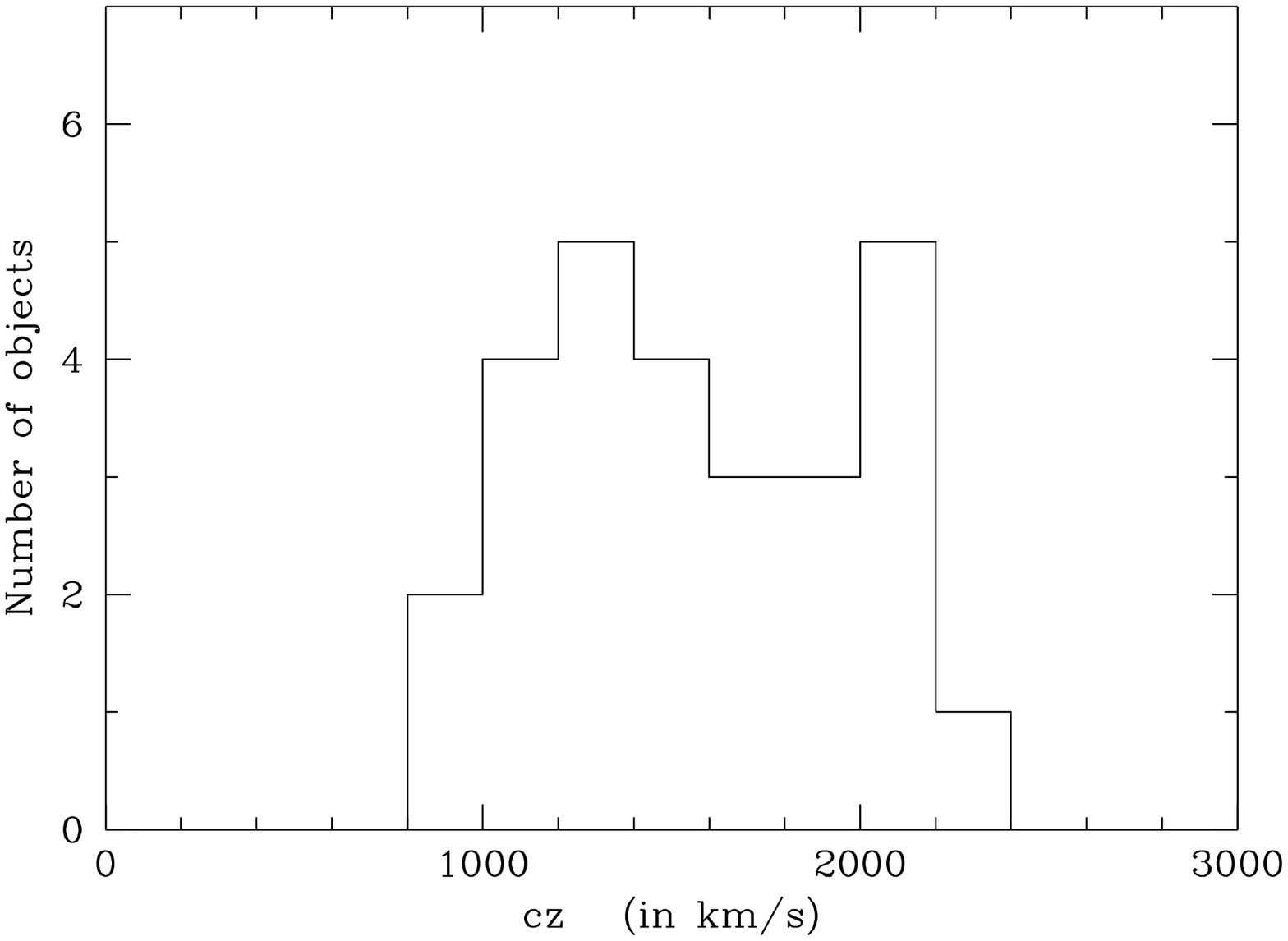,width=\one_wide}
\centering
\caption{ The heliocentric radial velocity distribution of Fornax Cluster 
galaxies from the {\em FSS} in Field~1. 
Objects with velocities $< 600$~kms$^{-1}$ 
(Galactic stars) are not shown for the sake of clarity. 
\label{fig_is_clusterfss}
}
\vspace*{-16mm}
\psfig{file=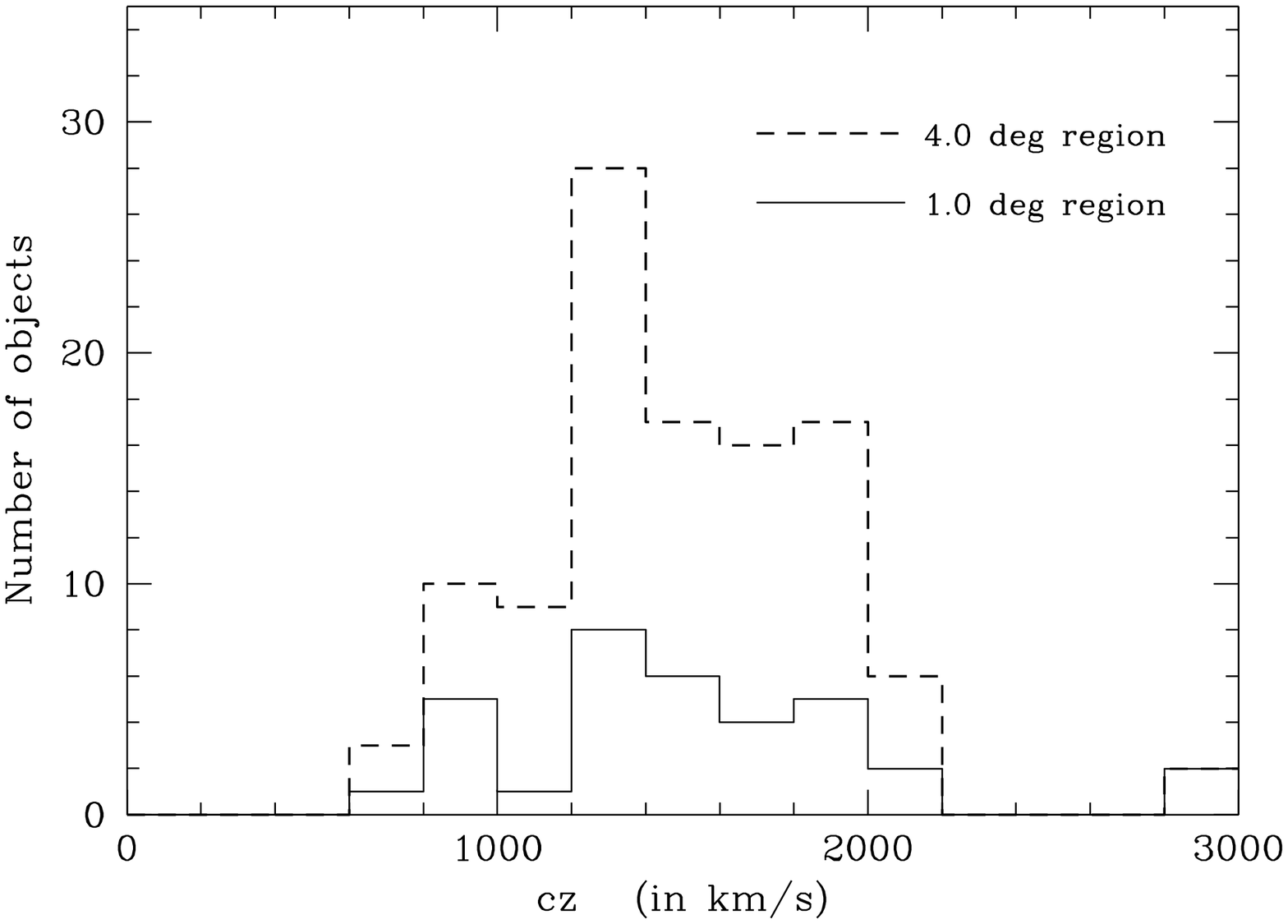,width=\one_wide}
\centering
\caption{ The heliocentric radial velocity distribution of Fornax Cluster 
galaxies compiled from the {\em NASA Extragalactic Database} 
within {\it FSS} Field~1 (solid histogram) and within a 4.0-degree radius 
region centred on the field (dashed histogram). 
The area of the 4.0-degree radius region is, of course, 16 times that 
of Field~1. 
\label{fig_is_clusterned}
}
\end{figure}

Figure~\ref{fig_is_clusterfss} shows the velocity distribution 
of Fornax Cluster galaxies from the {\em FSS}. 
The mean heliocentric radial velocity from the {\em FSS} data is 
$1560 \pm 80\:\mbox{km}\:\mbox{s}^{-1}$ (26 galaxies). 
This compares with $1540 \pm 50\:\mbox{km}\:\mbox{s}^{-1}$
from Jones \& Jones. Recall
that the Jones \& Jones galaxies are much brighter than ours 
(roughly $-21 \leq M_{\rm B} \leq -15$ as against 
$-14 \leq M_{\rm B} \leq -11$)
and are spread over a much larger area, 6 degrees or about
1.6~Mpc across compared to our 2~degrees or 0.5~Mpc. A velocity dispersion
can be estimated fairly unambiguously as there are no galaxies with
velocities less than 900 or between 2300 and $3000\:\mbox{km}\:\mbox{s}^{-1}$. 
Our 26 galaxies give an observed radial velocity dispersion of 
$380 \pm 50\:\mbox{km}\:\mbox{s}^{-1}$, compared to the 
$391\:\mbox{km}\:\mbox{s}^{-1}$ of Jones \& Jones. 

The {\em FSS} velocity distribution can also be compared with the
equivalent distribution compiled from all published redshift
data. Figure~\ref{fig_is_clusterned} presents the velocity data from
{\em NED}. These give a mean heliocentric radial velocity of $1450 \pm
70\:\mbox{km}\:\mbox{s}^{-1}$ (32 galaxies), and a velocity dispersion
of $370 \pm 50\:\mbox{km}\:\mbox{s}^{-1}$, entirely consistent with
the {\em FSS} results. The {\em NED} results generally, though not
entirely, apply to the brighter cluster galaxies.

     Fornax is an apparently well relaxed, regular cluster as judged
by its central density concentration and low spiral content.  It would
require a very much larger sample of redshifts over the other fields
in order to explore properly any dynamical differences between
different galaxy populations. These initial results do not reveal any
difference in the dynamics between the bright and faint (giant and
dwarf) members of the cluster, although a wide-field study of brighter
galaxies (Drinkwater et al.\ 2000a) does suggest such a difference.

\subsection{The Velocity Structure behind the Fornax Cluster}
\label{sec_is_forncl}

\begin{figure}
\psfig{file=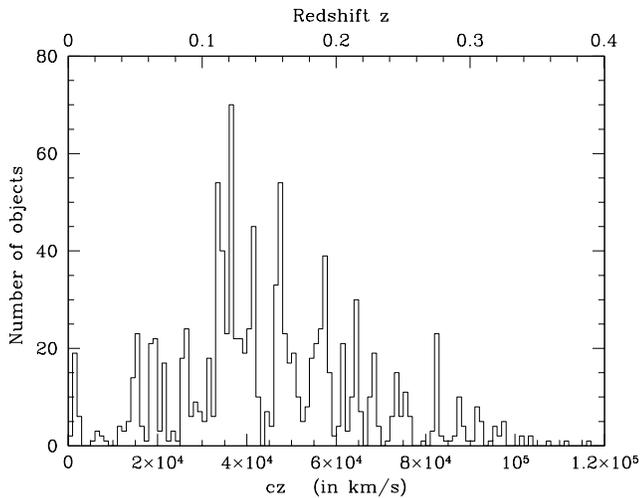,width=\one_wide}
\centering
\caption{ 
The radial velocity distribution from the {\em FSS} in Field~1 
showing the population behind the Fornax Cluster. 
Objects with velocities $< 600$~kms$^{-1}$ 
(Galactic stars) are not shown for clarity. 
\label{fig_is_backgndfss}
}
\end{figure}
\begin{figure}
\psfig{file=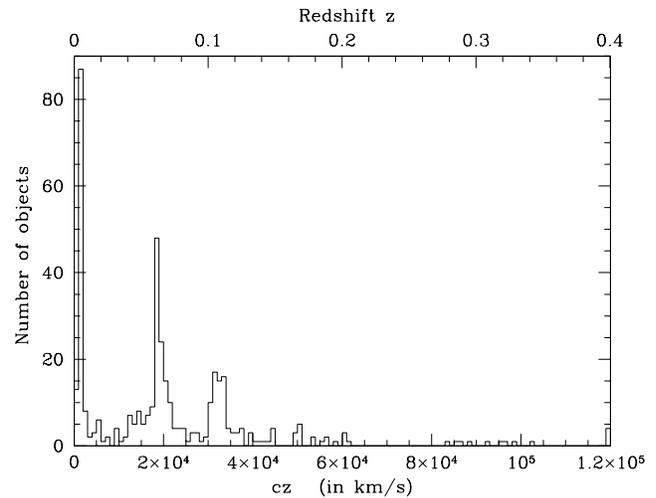,width=\one_wide}
\centering
\caption{ 
The radial velocity distribution from the 
{\em NASA Extragalactic Database} 
showing the population behind the Fornax Cluster. 
The figure shows all galaxies found in the {\em NED} within a 4.0~deg 
radius region centred on {\em FSS} Field~1. 
\label{fig_is_backgndned}
}
\end{figure}

Figure~\ref{fig_is_backgndfss} shows our redshift distribution behind 
the cluster. 
Immediately beyond the cluster,
as noted by Jones \& Jones (1980) and Phillipps \& Davies (1992), there
is a large void, extending some 40 Mpc (from the cluster mean redshift
to about $5000\:\mbox{km}\:\mbox{s}^{-1}$ assuming 
$H_0 = 75\:\mbox{km}\:\mbox{s}^{-1}\:\mbox{Mpc}^{-1}$). 
Beyond this ``Fornax Void'', we see the ubiquitous
`spiky' distribution (Broadhurst et al. 1990) showing more 
distant walls and filaments. 

Figure~\ref{fig_is_backgndned} shows the distribution of background 
galaxies taken from the {\em NED}. The difference in depth between 
the two data sets is immediately apparent: the {\FSS} results 
probe to much greater distances on account of the fainter magnitudes 
of the galaxies. Nevertheless, the first two main features in our 
distribution  clearly match the two peaks seen in the {\em NED} 
data (i.e. in the brighter galaxies).


\begin{figure*}
\vspace*{-30mm}
\psfig{file=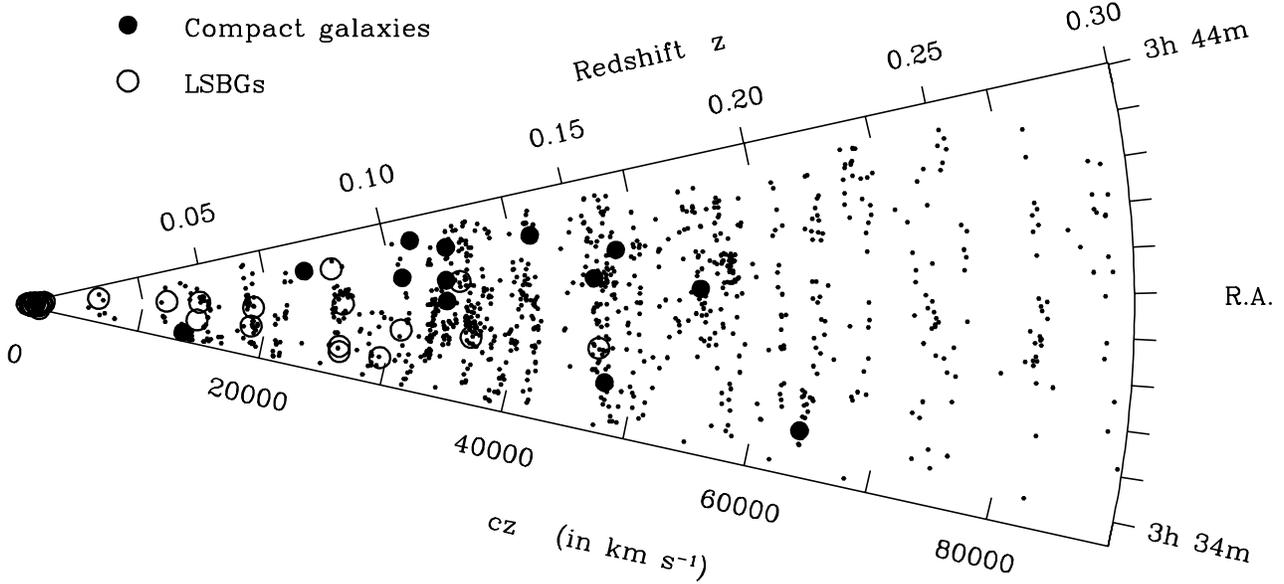,width=180mm}
\vspace*{-25mm}
\centering
\caption{ 
The radial velocity distribution of the galaxies 
behind the Fornax Cluster from Field~1 of the {\em FSS} as 
a function of J2000 right ascension showing the distribution of 
low surface brightness and compact galaxies. 
LSBGs, having intrinsic central surface brightnesses 
$> 22.5\; b_j \:\mbox{mag}\:\mbox{arcsec}^{-2}$, 
are shown as open circles. The compact galaxies of 
Drinkwater et al. (1999a, Paper~II) are shown as filled circles. 
Note that the right ascension axis has been greatly expanded (by a 
factor of 12.5 times) for 
clarity (falsely giving the peaks in the galaxy density the 
appearance of shells). 
\label{fig_is_conefssra2}
}
\end{figure*}


A standard cone diagram is shown in Figure~\ref{fig_is_conefssra2}, 
illustrating the skeleton of the large scale 3-D structure beyond Fornax. 
The median redshift of the entire galaxy sample is 0.15. 
This compares with a mean of 0.11 in the preliminary data from the 
2dF Galaxy Redshift Survey (Colless 1999). 
The data continue to map structure out to $z \simeq 0.30$, where there 
are still significant numbers of galaxies. 
The cluster J1556.15BL identified by Couch et al. (1991) lies in 
Field~1 at $z = 0.457$, but the density of {\em FSS} galaxies at 
this redshift is too small to show the cluster. 

\begin{figure}
\psfig{file=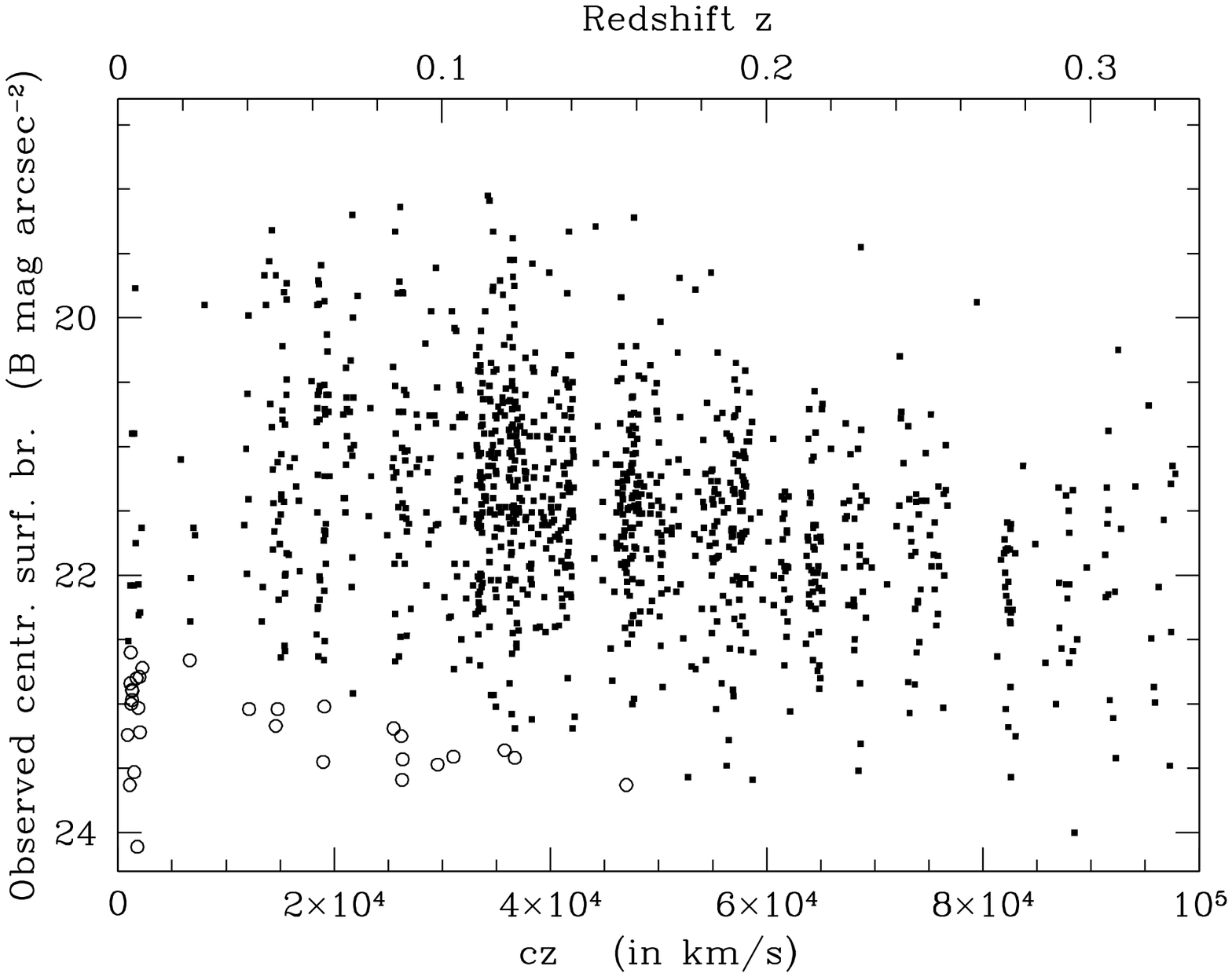,width=\one_wide}
\centering
\caption{ 
The central surface brightness of the {\em FSS}\ sample plotted 
against redshift. The blue surface brightnesses are taken 
from the surface photometry described in 
Section~\ref{ssec_resolvedobjs}. 
Low surface brightness galaxies, having intrinsic central surface 
brightnesses $> 22.5\; b_j \:\mbox{mag}\:\mbox{arcsec}^{-2}$,
are plotted as open circles. The conversion from observed to 
intrinsic surface brightnesses uses cosmological surface brightness 
dimming and $k$-corrections from Coleman, Wu and Weedman (1980). 
\label{fig_is_zsb}
}
\end{figure}

In addition to the general galaxy population, 
Figure~\ref{fig_is_conefssra2} also shows (as large solid points) the compact galaxies discussed in Paper~II. 
These objects have star-like images on Schmidt survey plates but 
the 2dF spectroscopy showed them to be compact star-forming galaxies 
at redshifts $0.04-0.21$. The figure also shows low surface brightness 
galaxies having intrinsic (cosmologically corrected) central surface 
brightnesses fainter than $22.5\; b_j \:\mbox{mag}\:\mbox{arcsec}^{-2}$, 
plotted as open circles. Despite their low surface brightnesses, these 
objects are sufficiently distant that they are too luminous to be dwarfs 
given the apparent magnitude limits of the survey (they have 
$M_{\rm B} \simeq -17$ to $-19$ for 
$H_0 = 75\:\mbox{km}\:\mbox{s}^{-1}\:\mbox{Mpc}^{-1}$).

Many authors (e.g. Phillipps \& Shanks 1987; Eder et al. 1989; 
Thuan et al. 1991; Loveday et al. 1995; Mo, McGaugh \& Bothun 1994) 
have discussed whether or not 
low luminosity and/or low surface brightness galaxies follow the same 
structures as the brighter component. Although, as stated earlier,
this sample is not yet complete -- so we can not use strictly objective
measures such as the galaxy correlation function (Phillipps \& Shanks 1987) --
we do have enough information in our distribution to see that the low
surface brightness galaxies (shown 
as open circles in Figures~\ref{fig_is_zsb} and \ref{fig_is_conefssra2}) 
do trace the same large scale structure and are not seen ``filling the 
voids''(Dekel \& Silk 1986). 
The present data extend this comparison of the distribution of LSBG 
with that of normal galaxies to significantly lower surface 
brightnesses than most other studies (or, indeed, will be possible
with the standard SLOAN or GRS samples). 
Similarly, the compact (high surface brightness) galaxies, 
which are also likely to be missing from other surveys, 
again follow the same overall large scale structure 
in Figure~\ref{fig_is_conefssra2} as the general galaxy 
population. This is unlike the suggestions from some previous 
emission line galaxy surveys (e.g. Salzer 1989) that
such objects can appear in very low density regions.


\section{Summary}
\label{sec_summary}

In this paper we have presented an overview of the \fss, the first
complete, all-object spectroscopic survey to cover a large area of
sky. This project has only been made possible by the advent of the
400-fibre Two-degree Field spectrograph on the Anglo-Australian
Telescope. In total we hope to observe some 14,000 objects to a 
magnitude limit of
\bj=19.7 --- both `stars' and `galaxies' --- in a 12\degsq\ area of sky
centred on the Fornax Cluster.

The main technical challenges of the project concern the preparation
of the target catalogue and the analysis of the resulting spectra. Our
input catalogues are based on UK Schmidt Sky Survey plates digitised
by the APM facility. We have demonstrated that the APM image
catalogues provide sufficiently accurate target positions and
photometry for the unresolved sources. For the resolved sources our
photometry is derived by fitting exponential profiles to the image
parameters measured by the APM. We have tested our calibration with
new CCD observations. We use a semi-automated procedure to classify
our spectra and measure radial velocities based on cross-correlation
comparison with a set of stellar spectra, two emission-line galaxy
spectra and one QSO spectrum. This procedure successfully identifies
stars, galaxies and QSOs completely independently of their image
morphology.

When the \fss\ is complete we will have a unique, complete, sample of
Galactic stars, Fornax Cluster galaxies, field galaxies and distant
AGN. We have discussed some of the scientific questions that can be
addressed with such a sample. The principal objective is to obtain 
an unbiased sample of cluster members, which includes compact galaxies 
and low surface brightness dwarfs, independent of a membership 
classification based on morphological appearance. 

Redshift/velocity distributions are presented here based on 
spectroscopic results from the first of four 2dF fields. 
The velocity distribution of Galactic stars can be 
understood in terms of a conventional three-component model of the Galaxy. 
The Fornax Cluster dwarf galaxies in the first 2dF field have 
a mean heliocentric 
radial velocity of $1560 \pm 80\:\mbox{km}\:\mbox{s}^{-1}$ and 
a radial velocity dispersion of $380 \pm 50\:\mbox{km}\:\mbox{s}^{-1}$. 
The Fornax Cluster is well-defined dynamically, with a low 
density of galaxies in the foreground and immediate background. 
Beyond $5000 \:\mbox{km}\:\mbox{s}^{-1}$, 
the large-scale structure behind the Fornax Cluster is clearly 
delineated out to a redshift $z \simeq 0.30$. 
The compact galaxies found behind the cluster by Drinkwater et al. 
(1999a) are found to follow the structures delineated by the general 
galaxy population, as are background low surface brightness galaxies. 
Some more detailed initial results have already been
presented elsewhere (Drinkwater et al.\ 1999a, 1999b).


\begin{acknowledgements}

This project would not be possible without the superb 2dF facility
provided by the AAO and the generous allocations of observing time we
have received from PATT and ATAC.  MJD is grateful for travel support
from the University of Bristol and the International Astronomical
Union. SP acknowledges the support of the Royal Society via a
University Research Fellowship. JBJ and JHD are supported by the UK
PPARC. SP, JBJ and RMS acknowledge the hospitality of the School of 
Physics, University of New South Wales.  Part of this work was
done at the Institute of Geophysics and Planetary Physics, under the
auspices of the U.S. Department of Energy by Lawrence Livermore
National Laboratory under contract No.~W-7405-Eng-48.

\end{acknowledgements}



\begin{thebibliography}{}

\bibitem[1989]{abell89} 
    Abell G.O., Corwin H.G. JR., Olowin R.P., 1989, ApJS, 70, 1

\bibitem[1993]{barden93}
    Barden S.C., Elston R., Armandroff T., \& Pryor C.P., 1993, in: Fiber
    Optics in Astronomy II, ASP Conf. Ser. 37, ed. P.M.Gray, p.223 

\bibitem[1982]{blair82} 
    Blair M., Gilmore G., 1982, PASP, 94, 742

\bibitem[1999]{bock99} 
    Bock D., Large M.I., Sadler E.M., 1999, AJ, in press, astro-ph/9812083

\bibitem[1987]{bothun87} 
    Bothun G.D., Impey C.D., Malin D.F., Mould J.R., 1987, AJ, 94, 23

\bibitem[1995]{boyce95} 
    Boyce P.J., Phillipps S., 1995, A\&A, 296, 26 

\bibitem[1991]{boyle91} 
    Boyle B.J., Jones L.R., Shanks T., 1991, MNRAS, 251, 482

\bibitem[1998]{boyle98} 
    Boyle B.J., Smith R.J., Shanks T., Croom S.M., Miller L., 
    Read M., 1997, 
    in: Cosmological Parameters and Evolution of the Universe, 
    Proc. IAU Symp. 183, ed.\ K. Sato, p. 6 

\bibitem[1990]{broadhurst90}
    Broadhurst T.J., Ellis R.S., Koo D.C., Szalay A.S., 1990, Nature, 
    343, 726

\bibitem[1984]{bunclark84} 
    Bunclark P., Irwin M.J., 1984, 
    in: Astronomy with Schmidt-type Telescopes, Proc. IAU Colloq. 78, 
    ed.\ M. Capaccioli, Reidel, Dordrecht, p. 147

\bibitem[1996]{bureau96} 
    Bureau M., Mould J.R., Staveley-Smith L., 1996, ApJ, 463, 60

\bibitem[1987]{caldwell87} 
    Caldwell N., Bothun G.D., 1987, AJ, 94, 1126

\bibitem[1988]{carney88} 
    Carney B.W., Latham D.W., Laird J.B., 1988, AJ, 96, 560 

\bibitem[1987]{cawson87} 
    Cawson M.G.M., Kibblewhite E.J., Disney M.J. and Phillipps S., 1987, 
    MNRAS, 224, 557

\bibitem[1990]{coleman80} 
    Coleman G.D., Wu C.-C., Weedman D.W. 1980, ApJSS, 43, 393

\bibitem[1999]{colless99} 
    Colless M.M., 1999, Phil. Trans. R. Soc. London, Ser. A,  357, 105 

\bibitem[1993]{colless93} 
    Colless M.M., Ellis R.S., Broadhurst T.J., Taylor K., 
    Peterson B.A., 1993, MNRAS, 261, 19

\bibitem[1984]{condon84} 
    Condon J.J., 1984, ApJ 287, 461

\bibitem[1992]{condon92} 
    Condon J.J., 1992, ARA\&A 30, 55 

\bibitem[1998]{condon98} 
    Condon J.J., Cotton W.D., Greisen E.R., Yin Q.F., Perley R.A., 
    Taylor G.B., Broderick J.J., 1998, AJ, 115, 1693. 

\bibitem[1998]{couch91} 
    Couch W.J., Ellis R.S., Malin D.F., MacLaren I., 1991, 
    MNRAS, 249, 606


\bibitem[1991]{croswell91}
    Croswell K., Latham D.W., Carney B.W., Schuster W.J., Aguilar L., 1991, 
    AJ, 101, 2078 

\bibitem[1998]{dalcanton98} 
    Dalcanton J.J., 1998, ApJ, 495, 251 

\bibitem[1988]{daviesetal88} 
    Davies J.I., Phillipps S., Cawson M.G.M., Disney M.J., 
    Kibblewhite E.J., 1988, MNRAS, 232, 239

\bibitem[1986]{dekel86} 
    Dekel A., Silk J., 1986, ApJ, 303, 39

\bibitem[1990]{disney90} 
    Disney M.J., Phillipps S., Davies J.I., Cawson M.G.M., 
    Kibblewhite E.J., 1990, MNRAS, 245, 175

\bibitem[1995]{drinkwater95} 
    Drinkwater M.J., Barnes D.G., Ellison S.L., 1995, PASA, 12, 248

\bibitem[1996]{drinkwater96}
    Drinkwater M.J., Barot J., Irwin M., 1996, in The Anglo-Australian 
    Observatory Newsletter, No.  79, p. 7

\bibitem[1998]{drinkwater98}
    Drinkwater M.J., Gregg M.D., 1998, MNRAS, 296, L15

\bibitem[2000a]{drinkwater00a}
    Drinkwater M.J., Gregg M.D., Holman B.A., Brown M., 2000a, 
    MNRAS, submitted

\bibitem[2000b]{drinkwater00b}
    Drinkwater M.J., Gregg M.D., Jones J.B., Phillipps S., 2000b, PASA,
    submitted (Paper~IV)

\bibitem[1999a]{drinkwater99a}
    Drinkwater M.J., Phillipps S., Gregg M.D., Parker Q.A., 
    Smith R.M., Davies J.I., Jones J.B., Sadler E.M., 1999a, 
    ApJ, 511, L97 (Paper~II)

\bibitem[1999b]{drinkwater99b}
    Drinkwater M.J., Phillipps S., Jones J.B., 1999b, 
    in: The Low Surface Brightness Universe, Proc. IAU Colloq. 171, 
    eds.\ J.I. Davies, C.D. Impey, S. Phillipps, 
    Astron. Soc. Pacific, San Francisco, p. 120

\bibitem[1989]{eder89}
    Eder J.A., Oemler A., Schombert J.M., Dekel A, 1989, ApJ, 340, 29

\bibitem[1989]{ferguson89} 
    Ferguson H.C., 1989, AJ, 98, 367 (FCC)

\bibitem[1988]{ferguson88} 
    Ferguson H.C., Sandage A., 1988, AJ, 96, 1520

\bibitem[1999]{folkes99} 
    Folkes S., Ronen S., Price I., et al., 1999, 
    MNRAS, 308, 459

\bibitem[1991]{francis91} 
    Francis P.J., Hewett P.C., Foltz C.B., Chaffee F.H., Weymann R.J., 
    Morris S.L., 1991, ApJ, 373, 465

\bibitem[1987]{freeman87}
    Freeman K.C., 1987, ARA\&A, 25, 603

\bibitem[1999]{freeman99} 
    Freeman K., 1999, 
    in: The Low Surface Brightness Universe, Proc. IAU Colloq. 171, 
    eds.\ J.I. Davies, C.D. Impey, S. Phillipps, 
    Astron. Soc. Pacific, San Francisco, p. 3 

\bibitem[1983]{gilmore83} 
    Gilmore G.F., Reid I.N., 1983, MNRAS, 202, 1025 

\bibitem[1989]{gilmore89} 
    Gilmore G.F., Wyse R.F.G., Kuijken K., 1989, ARA\&A, 27, 555

\bibitem[1995]{gunn95} 
    Gunn J., 1995, AAS, 186, 44.05

\bibitem[1999]{hilker99}
    Hilker M., Infante L., Vieira G., Kissler-Patig M., Richtler T., 1999, 
    A\&AS, 134, 75

\bibitem[1990]{irwin90} 
    Irwin M.J., Davies J.I., Disney M.J., Phillipps S., 1990, MNRAS, 245, 289

\bibitem[1994]{irwin1994} 
    Irwin M.J., Maddox S., McMahon R., 1994, Spectrum, 2, 14

\bibitem[1984]{jacoby84} 
    Jacoby G.H., Hunter D.A., Christian C.A., 1984, ApJSup, 56, 257

\bibitem[1980]{jones80} 
    Jones J.E., Jones B.J.T., 1980, MNRAS, 191, 685

\bibitem[1996]{dejong96}
    de Jong R.S., 1996, A\&A, 313, 45

\bibitem[1988]{killeen88}
    Killeen N.E.B., Bicknell G.V., Ekers R.D., 1988, ApJ, 325, 180 

\bibitem[1985]{kron85}
    Kron R.G., Koo D.C., Windhorst R.A., 1985, A\&A, 146, 38

\bibitem[1987]{vanderkruit87}
    van der Kruit P.C., 1987, A\&A, 173, 59 

\bibitem[1998]{kuijken89}
    Kuijken K., Gilmore G.F., 1989, MNRAS, 239, 605. 

\bibitem[1998]{kurtz98}
    Kurtz M.J., Mink D.J., 1998, PASP, 110, 934

\bibitem[1988]{lasker88} 
    Lasker B.M., Sturch C.R., Lopez C., Mallama A.D., McLaughlin S.F., 
    Russell J.L., Wisniewski W.Z., Gillespie B.A., Jenkner H., 
    Siciliano E.D., Kenny D., Baumert J.H., Goldberg A.M., Henry G.W., 
    Kemper E., Siegel M.J., 1988, ApJS, 68, 1

\bibitem[1998]{lewis98} 
    Lewis I.J., Glazebrook K., Taylor K., 1998, SPIE, 3355, 828

\bibitem[1995]{loveday95} 
    Loveday J., Maddox S.J., Efstathiou G., Peterson B.A., 1995, 
    ApJ, 442, 457

\bibitem[1998]{loveday98} 
    Loveday J., Pier J., 1998, 
    in: Wide Field Surveys in Cosmology, 14th IAP Colloquium, 
    eds.\ S. Colombi, Y. Mellier, B. Raban, Editions Fronti\`{e}res, 
    in press, astro-ph/9809179

\bibitem[1992]{majewski92} 
    Majewski S.R., 1992, ApJSS, 78, 87 

\bibitem[1993]{majewski93} 
    Majewski S.R., 1993, ARA\&A, 31, 575 

\bibitem[1996]{majewski96} 
    Majewski S.R., Munn J.A., Hawley S.L., 1996, ApJ, 459, L73

\bibitem[2000]{meyer00}
    Meyer M.J., Drinkwater M.J., Phillipps S., Couch W.J., 2000,
    MNRAS, submitted (Paper V) 

\bibitem[1995]{miller92} 
    Miller L. A., Cormack W., Paterson M., Beard S., Lawrence L., 1992.
    in: Digitised Optical Sky Surveys, 
    eds.\ H.T. MacGillivray, E.B Thomson, 
    Kluwer Academic Publishers, p. 133 

\bibitem[1994]{mo94}
    Mo H.J., McGaugh S.S. \& Bothun G.D., 1994, MNRAS, 267, 129

\bibitem[1999]{morshidiesslinger99} 
    Morshidi-Esslinger Z.B., Davies J.I. and Smith R.M., 1999, 
    MNRAS, 304, 297

\bibitem[1985]{morton85} 
    Morton D.C., Krug P.A., Tritton K.P., 1985, MNRAS, 212, 325

\bibitem[1994]{norris94} 
    Norris J.E., 1994, ApJ, 431, 645

\bibitem[1997]{phillipps97} 
    Phillipps S., 1997, 
    in: Wide-Field Spectroscopy, 
    eds.\ E. Kontizas, M. Kontizas, D.H. Morgan, G.P. Vettolani, 
    Kluwer Academic Publishers, p. 281

\bibitem[1992]{phillipps92} 
    Phillipps S., Davies J.I., 1992, 
    in: Digitised Optical Sky Surveys, 
    eds.\ H.T. MacGillivray, E.B Thomson, 
    Kluwer Academic Publishers, p. 295

\bibitem[1986]{phillipps85} 
    Phillipps S., Disney M.J., 1986, MNRAS, 221, 1039

\bibitem[1987]{phillipps87} 
    Phillipps S., Disney M.J., Kibblewhite E.J., Cawson M.G.M., 1987, 
    MNRAS, 229, 505

\bibitem[1995]{phillipps95}
    Phillipps S., Driver S.P., 1995, MNRAS, 274, 832

\bibitem[1998]{phillipps93} 
    Phillipps S., Parker Q.A., 1993, MNRAS, 265, 385

\bibitem[1998]{phillippsshanks87} 
    Phillipps S., Shanks T., 1987, MNRAS, 229, 621

\bibitem[1994]{roser94} 
    R\"{o}ser S., Bastian U., Kuzmin A., 1994, A\&AS, 105, 301

\bibitem[1993]{reid93}
    Reid N., Majewski S.R., 1993, ApJ, 406, 635

\bibitem[1989]{salzer89}
    Salzer J., 1989, ApJ, 347, 152

\bibitem[1985]{sandage85} 
    Sandage A., Binggeli B., Tammann G., 1985, AJ, 90, 385

\bibitem[1997]{smith97}
    Smith R.M., Driver S.P., Phillipps S., 1997, MNRAS, 287, 415

\bibitem[1998]{taylor98} 
    Taylor K., Cannon R.D., Parker Q.A., 1998,
    in: New Horizons from Multi-Wavelength Sky Surveys, Proc. IAU 
    Symposium 179, 
    eds.\ B.J. McLean, D.A. Golombek, J.J.E. Hayes, H.E. Payne, 
    Kluwer Academic Publishers, p. 135

\bibitem[1991]{thuan91} Thuan T.X., Alimi J.-M., Gott J.R., 
    Schneider S.E., 1991, ApJ, 370, 25

\bibitem[1979]{tonry79} Tonry J.L., Davis M., 1979, AJ, 84, 1511

\bibitem[1999]{Tresse99} Tresse L., Maddox S., Loveday J., Singleton C., 
    1999, MNRAS, 310, 262

\bibitem[1998]{Watson98}
    Watson F.G., Offer A.R., Lewis I.J., 1998, in: Fiber Optics in 
    Astronomy III, ASP Conf. Ser. 152, eds. S. Arribas, E. Mediavilla, 
    F.G. Watson, p. 50


\end{thebibliography}
\end{document}